\documentclass[12pt]{article}

\usepackage[margin=1in]{geometry} 
\usepackage{natbib}
\usepackage{appendix,soul}
\usepackage{graphicx}
\usepackage{tabularx,booktabs,array,dcolumn}
    \newcolumntype{L}[1]{>{\raggedright\let\newline\\\arraybackslash\hspace{0pt}}m{#1}}
    \newcolumntype{C}[1]{>{\centering\let\newline\\\arraybackslash\hspace{0pt}}m{#1}}
    \newcolumntype{R}[1]{>{\raggedleft\let\newline\\\arraybackslash\hspace{0pt}}m{#1}}
    \newcolumntype{d}[1]{D{.}{.}{0}}
\usepackage{caption,subcaption}
\usepackage{chngpage}
\usepackage{float,rotating}
\usepackage{color,xcolor}
\usepackage{amsmath,amsthm,amssymb,bbold,amsfonts,mathtools}
    \allowdisplaybreaks
\usepackage{setspace}
\usepackage[bottom]{footmisc}   
\usepackage{multirow}
\usepackage{marvosym}
\usepackage{verbatim}
\usepackage{dsfont}
\usepackage{titlesec}
    \titlespacing{\paragraph}{0pt}{1ex plus 0.5ex minus 0.1ex}{1em}
\usepackage[pdftex,
            pdfauthor={Pierre-Andre Chiappori, Alexandros Theloudis, Jorge Velilla, Jose Ignacio Gimenez-Nadal, Jose Alberto Molina},
            pdftitle={Commitment and the dynamics of household labor supply: new tests and evidence from Europe},
            pdfsubject={D12; D13; D15; J22; J31},
            pdfkeywords={Household behaviour; Intertemporal choice; Commitment; Collective model; Family labor supply; Dynamics; Wages; Europe; EU-SILC},
            hyperfootnotes=false,bookmarksopen=true]{hyperref}

\hypersetup{colorlinks=true,allcolors=blue}

\newcommand{\mcl}{\multicolumn}

\newcommand{\forceoddpage}{%
  \clearpage
  \ifodd\value{page}\else
    \hbox{}\thispagestyle{empty}\newpage
  \fi
}

\title{\textbf{Commitment and the\\dynamics of household labor supply:\\new tests and evidence from Europe}\thanks{ Chiappori: Department of Economics, Columbia University; \href{mailto:pc2167@columbia.edu}{pc2167@columbia.edu}. Theloudis: Department of Econometrics \& OR, Tilburg University; \href{mailto:a.theloudis@gmail.com}{a.theloudis@gmail.com}. Velilla: IEDIS, University of Zaragoza; \href{mailto:jvelilla@unizar.es}{jvelilla@unizar.es}. Gim\'enez-Nadal: IEDIS, University of Zaragoza; \href{mailto:ngimenez@unizar.es}{ngimenez@unizar.es}. Molina: IEDIS, University of Zaragoza; \href{mailto:jamolina@unizar.es}{jamolina@unizar.es}. We are grateful to Robert A. Pollak, who provided inspiration when writing this article, and to participants in the conference in his honor for helpful feedback. We also thank participants to the ERC Equiprice - Family Economics \& Matching conference at Sciences Po and especially Jean-Marc Robin for comments and suggestions. This work was supported by the Government of Aragón Project S32\_23R.}}

\author{
Pierre-Andr\'e Chiappori
\and
Alexandros Theloudis
\and
Jorge Velilla
\and
J. Ignacio Gim\'enez-Nadal
\and
Jos\'e Alberto Molina
}

\date{June 1, 2026}

\begin{document}

%%% Title
\maketitle

%%% Abstract 
\vspace*{-1.5ex}
\begin{abstract}
The ability of spouses to commit to future behavior has important implications for the allocation of resources between them and over time. Using a lifecycle collective model for household behavior, we propose new tests that distinguish between full, limited, and no commitment, based on the dynamic impact of wage shocks on household labor supply. A novelty of our approach is its ability to formally reject limited commitment, in addition to the other two types, exploiting sign restrictions from theory. We implement our tests across 15 European countries, drawing data from the EU-SILC over the years 2005--2019. We find that the elasticity of the Pareto weight with respect to favorable past wages is generally positive, consistent with bargaining under limited commitment. Past wage shocks thus induce bargaining effects on labor supply, empowering the recipient spouse and weakening the partner. Formally, we reject full and no commitment in all but 4 countries, but fail to reject limited commitment. 

\medskip


\medskip

\end{abstract}

%%%%%%%%%%%%%%%%%%%%%%%%%%%%%%%%%%%%%%%%
%%% Introduction
%%%%%%%%%%%%%%%%%%%%%%%%%%%%%%%%%%%%%%%%

\section{Introduction}

Traditionally, studies of household behavior revolve around the unitary model that takes the family as a single decision unit whose preferences are represented by one well-behaved utility function. This unitary approach has, over the last decades, come under criticism as its main predictions are typically rejected \citep[e.g.,][]{Lundberg1997, Duflo2003, PenglaseSozbir2026}. Several bargaining models appeared in the 1980s, recognizing that individual preferences may differ and modeling how these preferences interact.\footnote{See, e.g., \citet{Manser1980}, \citet{McElroy1981}, and \citet{Apps1986}.} \citet{Chiappori1988, Chiappori1992} proposed a general framework for household behavior, the collective model, where individuals are assumed to reach Pareto-efficient outcomes. Since then, several studies have proposed various extensions to the collective model\footnote{See \citet{Chiappori2022review} for a review of the vast empirical literature.}, including models based on a specific description of the cooperative decision process.\footnote{In particular, \citeauthor{LundbergPollak1993_Spheres}'s (\citeyear{LundbergPollak1993_Spheres}) `separate spheres' model uses cooperative bargaining with a specific definition of status quo points, which implies spouses restricting themselves to their traditionally gendered roles.}

In the real world, spouses interact repeatedly over the course of a marriage, sharing risk and offering each other intra-household insurance \citep{Blundell2016, Lise2019}. Yet, the early collective models were largely static, which has important limitations. In particular, they could not be used to evaluate policies with an intertemporal or risk-sharing dimension, as they abstracted from dynamic aspects of behavior. In practice, most of the literature on intertemporal behavior has long remained in the unitary field \citep[e.g.,][]{Scholz2006, Blundell2016, DeNardiFrenchJonesMcGee2025_WhySave}.

One reason for this absence is the complexity of dynamics in the collective model. In particular, the notion of \emph{commitment}, while crucial, raises specific difficulties. On the one hand, the ability of spouses to commit to future behavior is crucial because some commitment is necessary for sharing risk, for investing in children and common assets, or for enabling specialization at home. On the other hand, limits to commitment prevent the spouses from exploiting each other, offer a way out from a bad marriage, and restrict their ability to share risk. Understanding the extent of commitment in the household is therefore key for assessing risk sharing, intrahousehold inequality, and the impact of various policies on household outcomes. In pratice, any model involving repeated interactions must rely on specific assumptions regarding the agents' ability to commit to future actions \citep{Chiappori2017}. Predictions from the model will depend on the assumptions made and, as a result, policy recommendations will also vary with them. For example, a cash transfer to women will fail to empower women in certain commitment regimes, whereas in other regimes it will induce long-lasting effects.

Various ways of modeling commitment in a collective framework have been proposed. First best efficiency, for instance, requires spouses commit fully, at marriage, to a plan that disciplines all future allocations of resources between them, thereby ensuring ex ante efficient outcomes and full risk sharing. This full commitment assumption, for which \citet{LundbergPollak2003} coined the term BAMM (for Binding Agreements on the Marriage Market), has however been criticized as overly restrictive. In particular, such agreements may not be legally enforceable.\footnote{For instance, individuals cannot legally commit not to divorce their spouse.} \citet{LundbergPollak2003}, in contrast, analyze a situation in which spouses cannot commit at all. In this no commitment world, called BIM (for Bargaining In Marriage), the spouses renegotiate the plan in every period, which results in dynamic inefficiencies and prevents full risk sharing.\footnote{The BIM vs BAMM distinction has important consequences for modeling household formation. In particular, the standard Transferable Utility framework of \citet{Shapley1971} is relevant for BAMM, while BIM models tend to rely on Non Transferable Utility models \`a la \citet{Gale1962}. See \citet{Browning2014}.}

An interesting bridge between these two extreme situations is provided by the notion of limited commitment, introduced in family economics by \citet{Mazzocco2007}. In this framework, the spouses have outside options they cannot legally commit not to use, so continuation of the agreed plan must ensure that individual participation constraints are satisfied. If some shock to the state of the world results in a violation of one's participation constraint, then a renegotiation takes place with two possibilities: either a modification of the previous plan (if feasible without violating the spouse's constraint), or separation. Limited commitment ensures ex ante efficient allocations subject to participation constraints, thus providing second-best risk sharing \citep{Browning2014}.

The literature typically assumes a particular form of commitment. In contrast, we are interested in testing which commitment regime applies. \citet{Mazzocco2007} was the first to propose a test for full commitment, which he rejected using US consumption data. \citet{Blau2016} also rejected full commitment using inheritances in the US; \citet{Lise2019} rejected it using consumption and time use in Japan, and \citet{Toriyabe2025} rejected it using pension data, again in Japan. These studies exploit a common idea: full commitment implies constant Pareto weights over time, so they check whether current shocks affect the Pareto weight, a task facilitated by the fact that the weight is identified from consumption or labor supply behavior.

While these approaches reject full commitment, they are silent about the relevant non-full-commitment alternative. In this context, \citet{Theloudis2025} proposed a test for full and no commitment, strongly rejecting both using US labor supply data. Their intuition is that while current shocks separate full from non-full commitment \citep{Mazzocco2007}, past shocks matter in limited commitment because the Pareto weight exhibits memory, whereas they are irrelevant under no (and full) commitment. Hence, evidence that past shocks affect the Pareto weight provides grounds to reject no commitment as well.

None of these papers, however, directly test for limited commitment; instead, they offer only indirect evidence by ruling out alternative regimes. Moreover, evidence for Europe remains scarce despite the prevalence of family policies, such as parental leave, that have inherently intertemporal implications. This is precisely the gap this paper attempts to fill.

The present paper makes two contributions. First, we extend the existing tests by moving beyond testing for the presence of effects from current and past shocks to testing for the \emph{sign} of such effects. In no commitment, favorable current shocks empower the recipient spouse and weaken their partner. In limited commitment, favorable past shocks have analogous effects, though with magnitudes that decline over time. These asymmetric bargaining effects  translate into sign restrictions on behavior that sharpen the test for no commitment and can explicitly reject limited commitment. Suppose that, holding wealth constant, a good income shock earned by the wife reduces her leisure in the future. Existing tests would correctly conclude that both full and no commitment are therefore rejected. However, interpreting such findings as supporting the remaining option, i.e. limited commitment, would be incorrect. Standard bargaining theory requires that favorable shocks increase the wife's share of total income, thus her leisure by a standard income effect. Therefore, while having the same data requirements as its earlier counterparts, our test would also reject limited commitment in this case, precisely because the implied bargaining effect has the wrong sign.

Second, we test for commitment in Europe. While full commitment has been rejected in the US and Japan, it remains largely untested elsewhere. Europe is a particularly interesting case because several countries are currently revising family policies with inherently dynamic implications, whose effectiveness therefore depends on the degree of commitment. Using harmonized data from the European Union Statistics on Income and Living Conditions (EU-SILC) covering 15 countries, we estimate the dynamic effects of male and female wage shocks on spouses' labor supply and use these responses to characterize commitment in a comparable manner across different countries.

We find some heterogeneity across European countries, though the picture overall points towards limited commitment. We reject full and no commitment in all but 4 countries; we cannot reject full commitment in Bulgaria, Cyprus, Slovakia, and the UK, as shocks to wages show no significant effect on household labor supply. By contrast, we fail to reject limited commitment in all countries, perhaps with one marginal exception (Greece). In most cases, even in Slovakia and the UK for which we fail to formally reject full commitment, the elasticity of the Pareto weight with respect to own past wages is positive, implying bargaining effects that are economically consistent with limited commitment.

The paper relates to the newer strand of literature in labor, household, and development economics that studies the dynamics of household behavior, the determinants of women's bargaining power, and policies that shape behavior and power; \citet{Chiappori2017} and \citet{JayachandranVoena2025_WomensPower} provide excellent reviews. Most studies assume a particular form of commitment and obtain findings conditional on such assumption.\footnote{Some non-exhaustive examples: \citet{Voena2015} develops a limited commitment model to study unilateral divorce; \citet{Fernandez2017} do similarly under no commitment; \citet{Bronson2014} study college degree choices with limited commitment; \citet{Chiappori2018} study the implications of education choice with full commitment; \citet{Gousse2017} study home production without commitment; \citet{Low2018} study welfare reforms in limited commitment; \citet{Foerster2024} studies alimony under limited commitment; \citet{Reynoso2024} and \citet{Blasutto2025} study cohabitation and the marriage market under limited commitment; \citet{Bredemeier2023} estimate labor supply elasticities under limited commitment.} By contrast, we focus on testing for commitment, as also do, e.g., \citet{Mazzocco2007}, \citet{Lise2019}, and \citet{Theloudis2025} in the US and Japan.

The remainder of the paper is structured as follows. Section \ref{Section::Theory} presents the model and formulates the test for commitment. Section \ref{Section::Data} summarizes the data and describes the econometric strategy. Section \ref{Section::Results} shows the main results, while Section \ref{Section::Conclusion} concludes.

%%%%%%%%%%%%%%%%%%%%%%%%%%%%%%%%%%%%%%%%
%%% Theoretical framework
%%%%%%%%%%%%%%%%%%%%%%%%%%%%%%%%%%%%%%%%

\section{Theoretical framework}\label{Section::Theory}

%%% Setting and commitment regimes
\subsection{Setting and commitment regimes}\label{Section::Setting}

We follow a collective approach \citep{Chiappori1988, Chiappori1992}, which relies on cooperative game theory and assumes individuals reach Pareto-efficient outcomes. In particular, the dynamic framework is analogous to \citet{Theloudis2025}. Households consist of a male and a female spouse, denoted by $j=1,2$, who get married at time $t=0$ and live over a finite horizon of $T$ periods. Spouses enjoy utility from joint consumption ($q_t$) and joint leisure with the partner ($l_t$), and disutility from work hours ($h_{jt}$). The household problem can be expressed as
\begin{align}\label{Eq::Problem}
\begin{split}
\max\limits_{\{\mathbf{C}_t\}_{t=0}^T} ~ & ~ \sum\limits_{j} \mathds{E}_0 \sum\limits_{t} \beta^t  \mu_{j(t)} u_j (q_t, l_t, h_{jt}) \\
\text{s.t.} ~ & ~ (1+r) a_t + \tau(w_{1t}h_{1t} + w_{2t}h_{2t}) = q_t + a_{t+1}, ~~ \forall t,
\end{split}
\end{align}
where $\mathbf{C}_t = \{ q_t, l_t, h_{1t}, h_{2t}, a_{t+1}\}$ is the set of choice variables in period $t$, $a_t$ is common financial assets, $r$ is the deterministic interest rate, $\tau$ maps gross earnings $y_t = w_{1t}h_{1t} + w_{2t}h_{2t}$ into disposable income, and $w_{jt}$ is spouse $j$'s wage. In the objective function, $u_j$ is $j$'s utility over consumption and hours, which is continuously twice differentiable. The term $\mu_{j(t)}$ is the so-called Pareto weight, which summarizes the intra-household contract and represents how spouses allocate resources and power between them. As we show below, it varies with time under limited and no commitment, but is constant in full commitment -- hence subscript $t$ is in brackets. The household problem may be subject to additional constraints, in particular participation constraints under limited commitment, but those are captured by $\mu_{j(t)}$ so they can be put in the background here.\footnote{Joint leisure may require that spouses synchronize the timing of market work; we abstract from timing choices or timing constraints but refer to \citet{Cosaert2023} for a discussion.}

Under full commitment, the spouses commit at $t=0$ to a plan that disciplines their future actions, and outcomes are ex ante efficient. The Pareto weight is determined at marriage given some initial distribution factors $\mathbf{\Theta}_0$, a set of variables known or predicted at $t=0$, and remains constant over time, $\mu_{j(t)} \equiv \mu_{j} = \mu_{j} (\mathbf{\Theta}_0)$, as shown in \citet{Browning2014}. 

Under no commitment, the other extreme, the spouses do not commit to a future plan. New information revealed each period (e.g., wages, stochastic distribution factors $\mathbf{Z}_t$) changes the allocation of resources because the Pareto weight shifts reflecting the prevailing economic environment, $\mu_{j(t)} \equiv \mu_{jt} = \mu_{j} (\mathbf{\Theta}_0, w_{1t}, w_{2t}, \mathbf{Z}_t,a_t)$.\footnote{Distribution factors are exogenous variables that affect intra-household bargaining but not preferences or the budget constraint, conditional on household income \citep{Bourguignon2009}. $\mathbf{Z}_t$ represents distribution factors that vary during the course of the marriage, while $\mathbf{\Theta}_0$ represents variables that remain constant.} The spouses bargain repeatedly over the marital surplus, and outcomes are ex ante inefficient \citep{Lise2019}.

Under limited commitment, the spouses commit to future allocations up to the point that one's participation constraint is violated (i.e., one person is better off outside the household). Behavior is then second-best efficient, and program \eqref{Eq::Problem} is subject to individual rationality (participation) constraints that ensure that spouses enjoy their relationship at least as much as their outside option. As long as these constraints are satisfied, the spouses' bargaining power is summarized by the \emph{past} Pareto weight. Once a participation constraint is violated, the bargaining power of the constrained spouse must increase just enough for them to be indifferent between the relationship and the outside option.\footnote{We assume that this renegotiation is possible. Otherwise, the marriage comes to an end.} These dynamics are described by $\mu_{j(t)} \equiv \mu_{jt} = \mu_{jt-1} + \nu_{jt}(w_{1t}, w_{2t}, \mathbf{Z}_t, a_t, \mu_{jt-1})$, where $\nu_{jt}$ is the Lagrange multiplier on $j$'s participation constraint, and $\mu_{j0} = \mu_{j} (\mathbf{\Theta}_0)$ is the Pareto weight at marriage \citep{LigonThomasWorrall2002}. The initial Pareto weight will thus change if a participation constraint is violated.

\citet{Theloudis2025} show that these three dynamic planning problems share a common recursive form, subject to restrictions in the Pareto weight $\mu_{j(t)}$.\footnote{For completeness, we show these details in Appendix \ref{Appendix::Theory}.} They collect the period Pareto weights across commitment regimes as follows
\begin{equation}\label{Eq::Nesting}
\begin{array}{lcl}
\text{FC:}		& \mu_{j(t)} 	& = \mu_{j}(\mathbf{\Theta}_0) , \\
\text{NC:}	& \mu_{j(t)} 	& = \mu_{j}(\mathbf{\Theta}_0,w_{1t},w_{2t},\mathbf{Z}_t,a_{t}),  \\
\text{LC:} 	& \mu_{j(t)} 	& = \mu_{j}(w_{1t},w_{2t},\mathbf{Z}_t,a_{t},\mu_{jt-1}) \text{, and substituting $\mu_{jt-1}$ recursively} \\
			&			& = \mu_{j}(\mathbf{\Theta}_0,w_{1t},w_{2t},\mathbf{Z}_t,a_{t},w_{1t-1},w_{2t-1},\mathbf{Z}_{t-1},a_{t-1}, ~\dots~ ), 
\end{array}
\end{equation}
and show that the variables that enter the full commitment weight are nested within the set of variables that enter the no commitment weight, which in turn are nested within the set that enters the limited commitment weight. In other words, \emph{current} news to the household economic environment (i.e., $w_{1t}$, $w_{2t}$, $\mathbf{Z}_t$) do not matter in full commitment but do matter in the other alternatives. Besides that, \emph{past} news (i.e., $w_{1t-1}$, $w_{2t-1}$, $\mathbf{Z}_{t-1}$) that enter through $\mu_{jt-1}$ do not matter in full and no commitment but do matter in limited commitment. Older news from $t-2$ and earlier, i.e., the dots in the last line of \eqref{Eq::Nesting}, provide additional exclusion restrictions that separate limited commitment from the other regimes.

%%% Optimality conditions
\subsection{Optimality conditions}\label{Section::Equations}

\citet{Theloudis2025} describe alternative ways to estimate the optimality conditions that solve the household problem. We follow their quasi-reduced-form approach, deriving the intra-temporal optimality conditions for labor supply and carrying out a log-linearization of the marginal utility of hours as in \citet{Blundell2016}.\footnote{We focus on labor supply because assignable consumption is not observed empirically.} These operations, the details of which are in Appendix \ref{Appendix::Theory}, yield expressions for the growth rates of work hours of spouse $j=1,2$ at time $t$, given by 
\begin{align}\label{Eq::FOCs}
\begin{split}
    \Delta \log h_{jt} = \delta_{jt} h_{jt-1}^{-1} \times \Big\{ & 
    \underbrace{\Delta \log (1-\chi_{t})}_{\text{tax effects}} 
    - \underbrace{\kappa_{t} s_{-jt-1}\Delta \log y_{-jt}}_{\mathclap{\substack{\text{tax disincentives}\\\text{from partner earnings}}}}
    + \underbrace{\Delta \log \lambda_{t}}_{\mathclap{\substack{\text{wealth and}\\\text{income effects}}}} 
    - \underbrace{\zeta_{j}^q q_{t-1} \Delta \log q_{t}}_{\mathclap{\substack{\text{consumption}\\\text{complementarity}}}} \\
   &- \underbrace{\zeta_{j}^l l_{t-1} \Delta \log l_{t}}_{\mathclap{\substack{\text{joint leisure}\\\text{complementarity}}}}
    + \underbrace{(1-\kappa_{t} s_{jt-1}) \Delta \log w_{jt}}_{\mathclap{\substack{\text{substitution effects}}}} 
    - \underbrace{\Delta \log \mu_{jt}}_{\mathclap{\substack{\text{\emph{bargaining effects}}}}} \Big\}, 
\end{split}
\end{align}
where $\Delta$ is the first difference between $t-1$ and $t$, $-j$ indicates $j$'s partner, $s_{jt}\geq0$ is $j$'s earnings share, $y_{-jt}$ is the partner's earnings, $\lambda_t$ is the Lagrange multiplier on the budget constraint, and the terms $\chi_{t}$ and $\kappa_{t}$ represent the proportionality and progressivity of the tax and benefits system if we let $\tau(y_{t}) \approx (1-\chi_{t})y_{t}^{1-\kappa_{t}}$ as in \citet{Blundell2016}. Parameter $\delta_{jt} = (\alpha_{j}^{-1} + \kappa_{t} s_{jt-1} h_{jt-1}^{-1})^{-1}>0$, where $\alpha_{j}>0$ is $j$'s Frisch elasticity of labor supply scaled by his/her work hours, allows to sign the terms in \eqref{Eq::FOCs} and, in particular, allows to sign the \emph{bargaining effects}: if a given variable empowers (weakens) spouse $j$ increasing (decreasing) his/her Pareto weight, it will also decrease (increase) $j$'s hours of work. 

However, the (change in the) Pareto weight $\Delta \log \mu_{jt}$ is unobserved, which limits the empirical applicability of equation \eqref{Eq::FOCs}. We overcome this by log-linearizing the Pareto weight and substituting it backwards recursively. Consider the most general weight under limited commitment, $\mu_{jt} = \mu_j (w_{1t}, w_{2t}, z_{jt}, a_{t}, \mu_{jt-1})$, with $\mu_{j0} = \mu_j(\theta_{j0})$, where $z_{jt}\in \mathbf{Z}_t$ is one stochastic distribution factor and $\theta_{j0}\in \mathbf{\Theta}_0$ is one initial distribution factor, both assumed to empower spouse $j$. We introduce the full set of possible distribution factors, i.e., including also $z_{-jt}\in \mathbf{Z}_t$ and $\theta_{-j0}\in \mathbf{\Theta}_0$ that are meant to empower the partner $-j$, in Section \ref{Section::Test} subsequently. It follows that
\begin{align}\label{Eq::Approx_mu}
    \Delta\log \mu_{jt} &\approx 
    \underbrace{e_{[\mu_j, w_1]} \Delta\log w_{1t} + e_{[\mu_j, w_2]} \Delta\log w_{2t} + e_{[\mu_j, z_j]} \Delta\log z_{jt}}_{\text{news at $t$}} + e_{[\mu_j, a]} \Delta\log a_{t} \\
    \notag 
    &+ e_{[\mu_j, \mu_{jL}]}\big\{  \underbrace{e_{[\mu_j, w_1]} \Delta\log w_{1t-1} + e_{[\mu_j, w_2]} \Delta\log w_{2t-1} + e_{[\mu_j, z_j]} \Delta\log z_{jt-1}}_{\text{news at $t-1$}} + e_{[\mu_j, a]} \Delta\log a_{t-1} \big\} \\
    \notag 
    &+ e_{[\mu_j,\mu_{jL}]}^2 \big\{ \text{news at $t-2$} \big\} 
     + e_{[\mu_j,\mu_{jL}]}^2 e_{[\mu_j, a]} \Delta\log a_{t-2} + ~\dots~ 
     + e_{[\mu_j,\mu_{jL}]}^t e_{[\mu_j,\theta_j]} \theta_{j0},
\end{align}
where $e_{[\mu_j, \cdot]}$ is the elasticity of $\mu_j$ w.r.t each argument; $\mu_{jL}$ is short notation for $\mu_{jt-1}$. 

In non-full commitment, and assuming that positive changes in $w_{jt}$ and $z_{jt}$ empower $j$ and weaken $-j$, it follows that $e_{[\mu_j, w_j]}>0$, $e_{[\mu_j, z_j]}>0$, and $e_{[\mu_j, w_{-j}]}<0$. Because $\theta_{j0}$ also empowers $j$, it has to be $e_{[\mu_j, \theta_j]}>0$. Furthermore, the Pareto weight in limited commitment is persistent, i.e., it has memory, so it follows that $e_{[\mu_j, \mu_{jL}]}>0$. By contrast, the sign of $e_{[\mu_j, a]}$ is ambiguous and depends on which spouse benefits more from wealth as single. 

Assembling \eqref{Eq::FOCs} and \eqref{Eq::Approx_mu}, pooling common terms together, and simplifying the notation yields the quasi-reduced-form equation for hours growth of spouse $j$, namely
\begin{align}\label{Eq::Equation}
\begin{split}
\Delta \log h_{jt} = h_{jt-1}^{-1} & \times \Big\{ \text{controls for tax effects} + \text{controls for partner earnings} \\[-3pt]
    &+ \text{controls for wealth/income effects} \\[-3pt]
	&+ \text{controls for consumption \& leisure complementarities} \\[-3pt]
	&+ \sum\limits_{\tau=0}^{t-1}\beta_{j[\omega_{jt-\tau}]} \Delta\log w_{jt-\tau} + \sum\limits_{\tau=0}^{t-1} \beta_{j[\omega_{-jt-\tau}]} \Delta\log w_{-jt-\tau} \\
	&+ \sum\limits_{\tau=0}^{t-1} \beta_{j[z_{jt-\tau}]} \Delta\log z_{jt-\tau} + \sum\limits_{\tau=0}^{t-1} \beta_{j[a_{t-\tau}]} \Delta\log a_{t-\tau} + \beta_{j[\theta_j]} \theta_{j0} \Big\}.
\end{split}
\end{align}
This equation lies at the epicentre of our test for commitment.\footnote{A note on notation: the $\beta$'s are reduced-form coefficients that subsume the structural parameters that multiply each variable. This underlying structure is not important here but is made explicit in Appendix \ref{Appendix::Theory}.}

%%% Testing
\subsection{Testing for commitment}\label{Section::Test}

Our test exploits assignable bargaining effects, i.e., effects through signed changes in the Pareto weight, so our coefficients of interest are those accompanying wages and distribution factors. An exception is $\beta_{j[\omega_{jt}]}$, the coefficient on one's current wage, which captures both substitution and bargaining effects of own wages, so it cannot be used for testing.\footnote{Another exception is $\beta_{j[a_{t-\tau}]}$, $\tau\geq0$, the coefficient on current and past wealth, because wealth also induces wealth effects and the sign of the Pareto weight with respect to wealth is ambiguous.}\textsuperscript{,}\footnote{\citet{Theloudis2025} use $\beta_{j[\omega_{jt}]}$ and additional restrictions to structurally estimate $e_{[\mu_j, w_j]}$, the Pareto weight elasticity w.r.t. own wages. This elasticity can be used for testing but we abstract from this here.} By contrast, $\beta_{j[\omega_{-jt}]}$ reflects the bargaining effect from the partner's current wage, $\beta_{j[\omega_{jt-\tau}]}$ and $\beta_{j[\omega_{-jt-\tau}]}$, $\tau\geq1$, capture bargaining effects of past wages, $\beta_{j[z_{jt-\tau}]}$, $\tau\geq0$, capture bargaining effects from current and past distribution factors, and $\beta_{j[\theta_j]}$ reflects bargaining effects from the initial distribution factor, all of which enter exclusively through the Pareto weight. 

\citet{Theloudis2025} formalize this idea. They exploit the implications of \eqref{Eq::Nesting}, i.e., the nesting of variables that enter the Pareto weight in each commitment case, for the household labor supply equations \eqref{Eq::Equation}, and formulate testable hypotheses for full and no commitment:
\begin{equation*}%\label{Eq::Test}
\begin{array}{rcccccl}
{\cal H}_{0}^\text{FC}:    ~ &\beta_{j[\omega_{-jt}]}~= &\beta_{j[z_{kt}]}~=        &\beta_{j[\omega_{kt-\tau}]}~=      &\beta_{j[z_{kt-\tau}]}~=       &\beta_{j[\theta_k]}~=&0; \\[3pt]
{\cal H}_{0}^\text{NC}:    ~ &                          &                           &\beta_{j[\omega_{kt-\tau}]}~=      &\beta_{j[z_{kt-\tau}]}~=       &\beta_{j[\theta_k]}~=&0,
\end{array}
\end{equation*} 
for $j,k=1,2$ and $\tau=1,...,t-1$. These hypotheses concern the coefficients on the full set of stochastic and initial distribution factors, i.e., on wage shocks and on factors $z_{kt-\ell}\in \mathbf{Z}_{t-\ell}$ and $\theta_{k0}\in \mathbf{\Theta}_0$, for $k=1,2$ and $\ell=0,...,t-1$. The idea is that wages and distribution factors should not affect the Pareto weight in full commitment (the weight is constant over time), whereas \emph{past} wages and distribution factors should not affect it in no commitment (the weight changes with contemporaneous information but history does not matter).

However, these tests are based on the \emph{presence} (or lack thereof) of effects from current and past bargaining variables, while theory also disciplines the \emph{sign} of those effects. Favorable \emph{current} shocks received by $j$ improve his/her bargaining power under non-full commitment, decreasing his/her labor supply and increasing the partner's, while favorable \emph{past} shocks improved his/her bargaining power in the past, and through persistence in the Pareto weight under limited commitment, they also decrease his/her current labor supply and increase the partner's. These sign restrictions allow us to strengthen the test for no commitment \emph{and} explicitly test for limited commitment through two new hypotheses: 
\begin{equation*}%\label{Eq::TestExtended}
\begin{array}{rllll}
{\cal H}_{0}^\text{NC-new}: ~ &\beta_{j[\omega_{-jt}]}>0,       &\beta_{j[z_{jt}]}<0,               &\beta_{j[z_{-jt}]}>0,              &\\  
                              &\beta_{j[\omega_{kt-\tau}]}=0,   &\beta_{j[z_{kt-\tau}]}=0,          &\beta_{j[\theta_k]}=0;             &\\[7pt]
{\cal H}_{0}^\text{LC}:     ~ &\beta_{j[\omega_{-jt}]}>0,       &\beta_{j[z_{jt-\ell}]}<0,          &\beta_{j[z_{-jt-\ell}]}>0,         &\\   
                              &\beta_{j[\omega_{jt-\tau}]}<0,   &\beta_{j[\omega_{-jt-\tau}]}>0,    & \beta_{j[\theta_j]}<0,            &\beta_{j[\theta_{-j}]}>0,   
\end{array}
\end{equation*}
for $j,k=1,2$; $\ell=0,...,t-1$; and $\tau=1,...,t-1$.

%%%%%%%%%%%%%%%%%%%%%%%%%%%%%%%%%%%%%%%%
%%% Data
%%%%%%%%%%%%%%%%%%%%%%%%%%%%%%%%%%%%%%%%

\section{Data and empirical implementation}\label{Section::Data}

%%% Sample and variables
\subsection{Sample and variables} 

We use data from the longitudinal samples of the European Union Statistics on Income and Living Conditions (EU-SILC), over the period 2005–2019. The EU-SILC comprises comparable surveys across several European countries, conducted annually by Eurostat as part of the European Statistical System. It provides data at the family and individual level for surveyed households, and includes information on income, employment, poverty, and living conditions, among other factors. The longitudinal samples of the EU-SILC consist of rolling panels and surveyed households are followed for up to four consecutive years.

We select a sample of married and unmarried couples \citep{Chiappori2002}, aged between 21 and 65 years old \citep[i.e., similar to][]{Mazzocco2007,Theloudis2025}. We require complete data on key variables (demographics, earnings and wages, work hours, consumption, and wealth), and that both partners participate in the labor market.\footnote{Conditional on all other selection criteria, about 80\% of men and 70\% of women work in our sample.} Finally, the estimating equations require at least three consecutive periods of data, so we retain couples observed for at least three consecutive years.

The final sample consists of 105,413 observations (households $\times$ years), corresponding to 39,804 households across 15 countries: Austria, Belgium, Bulgaria, Cyprus, the Czech Republic, France, Greece, Hungary, Italy, Luxembourg, Poland, Romania, Slovakia, Spain, and the United Kingdom.\footnote{Croatia, Estonia, Ireland, Latvia, Lithuania, Malta, Portugal, Serbia, and Switzerland are omitted from the sample because of small sample sizes once we apply sample selection. Germany only appears in the data after 2018, while Denmark, Finland, Iceland, the Netherlands, Norway, Slovenia and Sweden have samples of persons, not of households, so we cannot include them in the sample because of missing key variables.} Because the estimating equations are in first difference, the effective number of observations entering our estimation is 65,716. Appendix \ref{Appendix::Sumstats} reports details on the sample selection and key variables, Table \ref{AppTable::Sample} shows the number of households and the survey years per country, and Table \ref{AppTable::Sumstatscc} provides summary statistics.

The main information needed to estimate equation \eqref{Eq::Equation} is the spouses' work hours and their current and past wages. The EU-SILC includes information on work hours per spouse and wage rates in \EUR/hour.\footnote{All monetary variables are expressed in 2019 prices.} We also need information on family earnings, expenditure, wealth, joint leisure, and demographics. Expenditure in the survey is defined as `usual' annual household expenditure. Wealth is not observed, so we proxy it by returns to wealth (interests, dividends, and profits from capital), assuming returns heterogeneity away. Joint leisure is also not observed; we proxy it by the partner's leisure, which is otherwise excluded from \eqref{Eq::Equation}. Appendix \ref{Appendix::Sumstats} provides further information on these choices.

%%% Estimating equations
\subsection{Estimating equations}

Equation \eqref{Eq::Equation} shows the labor supply equation for spouse $j=1,2$ that solves the household problem under the most general (limited commitment) Pareto weight. Some modeling choices are needed before we can bring this equation to the data.

First, we control for tax effects -- the term for $\chi_t$ in \eqref{Eq::FOCs} -- through individual and household demographics $\mathbf{x}_{jt}$.\footnote{$\mathbf{x}_{jt}$ includes interaction terms for year, year of birth, education, number of children and family members.} Second, we control for the partner's earnings by including $\Delta \log y_{-jt}$ in the equation, consistent with \eqref{Eq::FOCs}. Third, we control for wealth and income effects -- the term for $\lambda_t$ in \eqref{Eq::FOCs} -- by a polynomial in current and past growth rates and log levels of wealth and income. These controls subsume the wealth terms $\Delta\log a_{t-\tau}$ that enter via the Pareto weight, which is another reason why the bargaining effects of wealth cannot be used for testing. Fourth, we control for consumption and joint leisure by introducing terms for their respective growth rates, consistent with \eqref{Eq::FOCs}.

True time-varying distribution factors $z_{jt}$ are hard to find empirically, let alone in a comparable way across 15 countries \citep{Chiappori2017}. Fortunately, $z_{jt}$ is not needed to separate between the four commitment hypotheses; select current and past wage shocks serve as applicable distribution factors in this context, so we drop $z_{jt}$ completely. In practice, we do not observe households during their entire lifecycle. We thus consider shocks only from periods $t$ and $t-1$, which suffices for testing all hypotheses: time $t$ (partner's) shocks allow to separate full from non-full commitment, while $t-1$ shocks separate limited from no commitment; both periods serve to test for limited commitment. We include one initial distribution factor $\theta_j$, the age gap at marriage, assuming that a (much) younger spouse has a busier marriage market, hence more bargaining power at marriage.\footnote{We define $\theta_j$, the spouses' age gap at marriage, as a dummy which, when $j=1$, reflects the husband is younger and, when $j=2$, reflects the wife is 4 years younger \citep{Theloudis2025}.} 

We separate the deterministic component of wages, which agents typically anticipate and contract upon at marriage, from wage shocks $\omega_{jt}$, that are unanticipated and can thus induce bargaining effects in the course of the relationship, by assuming $\Delta\log w_{jt} = \boldsymbol{\pi}_j^{\prime} \mathbf{x}_{jt}^w + \omega_{jt}$. Here $\boldsymbol{\pi}_j^{\prime} \mathbf{x}_{jt}^w$ is the deterministic profile and $\mathbf{x}_{jt}^w \subseteq \mathbf{x}_{jt}$ is a vector of worker demographics \citep{Altonji2013}. These modeling choices yield the final equation for hours, $j=1,2$, given by
\begin{align}\label{Eq::EstimableEquation}
\begin{split}
\Delta \log h_{jt} = h_{jt-1}^{-1} 
    &\times \Big\{ b_{j[0]} + \mathbf{b}_{j[x]}^{\prime}\mathbf{x}_{jt} \\
	&+ \beta_{j[\omega_{jt}]} \omega_{jt} + \beta_{j[\omega_{-jt}]} \omega_{-jt} + \beta_{j[\omega_{jt-1}]} \omega_{jt-1} + \beta_{j[\omega_{-jt-1}]} \omega_{-jt-1} + \beta_{j[\theta_j]} \theta_{j0} \\ 
    &+ \text{controls for $y_{-j}$; wealth and income; $q$-$l$ complementarities} \Big\}.
\end{split}
\end{align}
Details on how we arrive at this final estimating equation are shown in Appendix \ref{Appendix::Theory}.

%%% Econometric approach
\subsection{Econometric approach}

We follow a two-stage approach to estimate the labor supply equation \eqref{Eq::EstimableEquation} per spouse. In the first stage, we regress wage growth $\Delta\log w_{jt}$ on observables $\mathbf{x}_{jt}^w$ (which coincide empirically with $\mathbf{x}_{jt}$), and obtain the wage shock $\omega_{jt}$ as the residual. We also regress hours growth $\Delta\log h_{jt}$ on observables $\mathbf{x}_{jt}$ interacted with inverse past hours to obtain residual hours net of cross-sectional heterogeneity, consistent with \eqref{Eq::EstimableEquation}. In the second stage, we regress residual hours on the remaining terms that reflect bargaining effects, income/wealth effects, and consumption and leisure complementarities. All equations are estimated using OLS, separately for each country. We use cluster-robust standard errors at the household level to account for heteroskedasticity and serial correlation within clusters \citep{Cameron2015}.

Our earlier modeling choices, in particular our choice to abstract from stochastic distribution factors and wage shocks older than $t-1$, give rise to the following testable hypotheses that we bring to the data:
\begin{equation}\label{Eq::Test_final}
\begin{array}{rlllll}
{\cal H}_{0}^\text{FC}:    ~ 	&\beta_{j[\omega_{-jt}]}=0,   & \beta_{j[\omega_{jt-1}]}=0, & \beta_{j[\omega_{-jt-1}]}=0, & \beta_{j[\theta_j]}=0, ~~& j\in\{ 1,2 \}; \\
{\cal H}_{0}^\text{NC}:    ~ 	&                             & \beta_{j[\omega_{jt-1}]}=0, & \beta_{j[\omega_{-jt-1}]}=0, & \beta_{j[\theta_j]}=0, & j\in\{ 1,2 \}; \\
{\cal H}_{0}^\text{NC-new}:	~   & \beta_{j[\omega_{-jt}]}>0,  & \beta_{j[\omega_{jt-1}]}=0, & \beta_{j[\omega_{-jt-1}]}=0, & \beta_{j[\theta_j]}=0, & j\in\{ 1,2 \}; \\
{\cal H}_{0}^\text{LC}:	~       & \beta_{j[\omega_{-jt}]}>0,  & \beta_{j[\omega_{jt-1}]}<0, & \beta_{j[\omega_{-jt-1}]}>0, & \beta_{j[\theta_j]}<0, & j\in\{ 1,2 \}.
\end{array}
\end{equation} 
${\cal H}_{0}^\text{FC}$ corresponds to the test for full commitment in \citet{Mazzocco2007}, cast here in terms of wage rather than income shocks, while ${\cal H}_{0}^\text{NC}$ corresponds to the test for no commitment proposed by \citet{Theloudis2025}. We assess these hypotheses here \emph{jointly} between wives and husbands, $j=1,2$, using a Wald statistic combining the estimated covariance matrices from the male and female equations into a single block-diagonal matrix.

Among the new hypotheses, ${\cal H}_{0}^\text{NC-new}$ combines equality and sign restrictions, whereas ${\cal H}_{0}^\text{LC}$ consists exclusively of sign restrictions. We implement both as likelihood-ratio tests that impose equality and inequality restrictions on regression coefficients \citep{Wolak1987,Wolak1989}. The test compares the fit of a constrained model (with the theoretical restrictions) to that of an unconstrained model. If the restrictions are correct, imposing them will not significantly reduce the likelihood; if one or more restrictions fail in the data, the constrained model's fit will be substantially worse. The resulting test statistic has a $\bar{\chi}^2$ distribution \citep{Gromping2010}. This approach differs from alternative methods, such as \citet{Chernozhukov2007} and \citet{Andrews2010}, which are appropriate in moment inequality settings. Since our hypotheses consist of equality and inequality restrictions on linear regression coefficients, \citet{Wolak1987, Wolak1989}'s tests are valid and considerably simpler.

%%%%%%%%%%%%%%%%%%%%%%%%%%%%%%%%%%%%%%%%
%%% Results
%%%%%%%%%%%%%%%%%%%%%%%%%%%%%%%%%%%%%%%%

\section{Results and extensions}\label{Section::Results}

%%% Reduced-form results
\subsection{Estimation results}

Table \ref{table::main} shows the main estimates from \eqref{Eq::EstimableEquation}, namely the coefficients on current wage shocks, past wage shocks, and the initial distribution factor, across 15 countries. Detailed results per country, including standard errors and the remaining parameters not shown here, appear in Appendix Tables \ref{table::detailsAT} to \ref{table::detailsUK}. Four main findings stand out. 

First, the coefficient on the \emph{partner's current} shock, $\beta_{j[\omega_{-jt}]}$, is positive whenever it is statistically significant. This is evidence against full commitment (which requires $\beta_{j[\omega_{-jt}]}=0$ as, conditional on partner's income, these shocks do not induce a reallocation) and in favor of limited and no commitment (which require $\beta_{j[\omega_{-jt}]}>0$ as these shocks empower the partner and lead to a reduction in own leisure and a rise in labor supply). Second, the coefficient on one's \emph{own past} shock, $\beta_{j[\omega_{jt-1}]}$, is negative whenever it is significant. This is evidence against full \emph{and} no commitment (which require $\beta_{j[\omega_{jt-1}]}=0$ as past shocks do not matter for current allocations) and in favor of limited commitment (which requires $\beta_{j[\omega_{jt-1}]}<0$ as these shocks empowered oneself in the past and the Pareto weight is persistent). Third, the coefficient on the \emph{partner's past} shock, $\beta_{j[\omega_{-jt-1}]}$, is mostly positive, providing further evidence for limited commitment. Fourth, the coefficient on the initial distribution factor, $\beta_{j[\theta_j]}$, shows mixed results but it is mostly negative, consistent with limited commitment. 

\begin{sidewaystable}[htp!]  
\begin{center}
\caption{Main parameter estimates}\label{table::main}
\resizebox{\textwidth}{!}{
\begin{tabular}{lcccccccccc}
\toprule
                            & Male            & Female          & Male            & Female          & Male            & Female          & Male            & Female          & Male            & Female          \\
                            & $(j=1)$         & $(j=2)$         & $(j=1)$         & $(j=2)$         & $(j=1)$         & $(j=2)$         & $(j=1)$         & $(j=2)$         & $(j=1)$         & $(j=2)$\\ 
\cmidrule{2-11}
                            & \mcl{2}{c}{Austria}       & \mcl{2}{c}{Belgium}       & \mcl{2}{c}{Bulgaria}      & \mcl{2}{c}{Cyprus}        & \mcl{2}{c}{Czech Rep.}    \\ 
\midrule
\multicolumn{11}{l}{\emph{Current shocks:}}\\
$\beta_{j[\omega_{jt}]}$    & $ -452.675$**   & $ -302.384$**   & $ -478.534$**   & $ -241.878$**   & $  -56.414$**   & $  -52.436$**   & $ -403.703$**   & $ -291.287$**   & $ -378.429$**   & $  -94.913$**   \\
$\beta_{j[\omega_{-jt}]}$   & $    9.747$     & $  -65.641$     & $  -56.030$     & $  -15.671$     & $   17.863$     & $   23.250$     & $   68.699$     & $    9.439$     & $  100.325$**   & $    9.562$     \\
\multicolumn{11}{l}{\emph{Past shocks:}}\\
$\beta_{j[\omega_{jt-1}]}$  & $  -71.976$**   & $  -38.797$**   & $  -64.177$**   & $   -9.725$     & $    1.720$     & $   13.438$     & $   -8.436$     & $    2.114$     & $    3.680$     & $  -86.776$**   \\
$\beta_{j[\omega_{-jt-1}]}$ & $   24.681$*    & $  -20.422$     & $   -0.252$     & $   31.876$     & $    0.827$     & $    4.393$     & $    0.715$     & $   47.612$*    & $    5.749$     & $   -0.642$     \\
\multicolumn{11}{l}{\emph{Initial distr. factor:}}\\
$ \beta_{j[\theta_j]}$        & $  -61.760$*    & $  -25.946$*    & $   -0.987$     & $   -4.098$     & $    6.855$     & $   -3.211$     & $   11.905$     & $    7.293$     & $   11.672$     & $  -14.947$     \\
\# obs.                     & $        3,170$ & $        3,170$ & $        4,115$ & $        4,115$ & $        3,482$ & $        3,482$ & $        3,541$ & $        3,541$ & $        6,069$ & $        6,069$ \\
\midrule\midrule
                            & \mcl{2}{c}{France}        & \mcl{2}{c}{Greece}        & \mcl{2}{c}{Hungary}       & \mcl{2}{c}{Italy}         & \mcl{2}{c}{Luxembourg}    \\ 
\midrule
\multicolumn{11}{l}{\emph{Current shocks:}}\\
$\beta_{j[\omega_{jt}]}$    & $ -667.063$**   & $ -170.289$*    & $ -542.965$**   & $ -431.505$**   & $ -193.258$**   & $ -297.211$**   & $ -386.510$**   & $ -483.033$**   & $ -399.093$**   & $ -242.285$**   \\
$\beta_{j[\omega_{-jt}]}$   & $   61.794$     & $   42.654$     & $  -42.414$     & $  -86.552$     & $   83.943$*    & $   17.680$     & $  -29.169$     & $  -35.228$     & $  -24.459$     & $  -55.821$     \\
\multicolumn{11}{l}{\emph{Past shocks:}}\\
$\beta_{j[\omega_{jt-1}]}$  & $  -21.885$     & $  -76.676$**   & $  -58.623$**   & $  -91.279$**   & $  -46.345$*    & $  -44.948$     & $  -49.644$**   & $  -59.728$**   & $  -65.318$*    & $  -56.195$**   \\
$\beta_{j[\omega_{-jt-1}]}$ & $  -36.392$*    & $  -55.618$     & $   -9.576$     & $   25.017$     & $   24.302$     & $   42.069$*    & $  -19.837$     & $  -18.546$     & $    6.640$     & $   42.829$     \\
\multicolumn{11}{l}{\emph{Initial distr. factor:}}\\
$ \beta_{j[\theta_j]}$        & $   13.192$     & $   -8.423$     & $   28.972$*    & $   -5.498$     & $   -7.415$     & $    1.293$     & $  -20.588$     & $    0.306$     & $    9.356$     & $  -17.283$     \\
\# obs.                     & $        2,292$ & $        2,292$ & $        3,095$ & $        3,095$ & $        2,721$ & $        2,721$ & $        6,779$ & $        6,779$ & $        3,768$ & $        3,768$ \\
\midrule\midrule
                            & \mcl{2}{c}{Poland}        & \mcl{2}{c}{Romania}       & \mcl{2}{c}{Slovakia}      & \mcl{2}{c}{Spain}         & \mcl{2}{c}{UK}            \\
\midrule
\multicolumn{11}{l}{\emph{Current shocks:}}\\
$\beta_{j[\omega_{jt}]}$    & $ -302.976$**   & $ -293.569$**   & $ -221.465$**   & $  -83.205$**   & $ -213.993$**   & $ -210.597$**   & $ -339.853$**   & $ -286.678$**   & $ -535.545$**   & $ -163.441$**   \\
$\beta_{j[\omega_{-jt}]}$   & $   17.773$     & $   65.818$     & $   88.348$**   & $   93.875$**   & $  -61.802$     & $   57.362$     & $  -29.861$     & $   41.527$     & $   32.395$     & $    1.784$     \\
\multicolumn{11}{l}{\emph{Past shocks:}}\\
$\beta_{j[\omega_{jt-1}]}$  & $    9.261$     & $  -75.948$**   & $  -52.199$*    & $  -87.949$**   & $  -11.121$     & $   -0.952$     & $   14.853$     & $  -57.977$*    & $  -46.425$     & $   -8.152$     \\
$\beta_{j[\omega_{-jt-1}]}$ & $    1.093$     & $  -19.034$     & $   33.043$     & $   24.572$     & $    3.090$     & $   24.163$     & $  -72.881$     & $   25.159$     & $   -0.302$     & $   15.219$     \\
\multicolumn{11}{l}{\emph{Initial distr. factor:}}\\
$ \beta_{j[\theta_j]}$        & $  -10.659$     & $   -2.326$     & $   12.958$     & $    2.314$     & $   10.155$     & $    4.841$     & $  -33.031$*    & $   -7.781$     & $   24.471$     & $   -9.798$     \\
\# obs.                     & $        8,351$ & $        8,351$ & $        4,912$ & $        4,912$ & $        4,440$ & $        4,440$ & $        5,136$ & $        5,136$ & $        3,845$ & $        3,845$ \\
\bottomrule
\end{tabular}}
\begin{minipage}{\textwidth}
\footnotesize
\textit{Notes}: The table presents the main estimates from \eqref{Eq::EstimableEquation} across 15 countries. Detailed results per country, including standard errors and the remaining parameters not shown here, appear in Appendix Tables \ref{table::detailsAT} to \ref{table::detailsUK}. $^{**}$ significant at the 1\%; $^{*}$ significant at the 5\%.
\end{minipage}
\end{center}
\end{sidewaystable}

In terms of countries, both male and female labor supply equations in Austria, Hungary, Italy, Luxembourg and Romania show evidence for limited commitment, as own past shocks relate negatively to current hours, with the coefficients being highly significant in all cases. Besides, the partner's current and past shocks relate positively to hours in several cases (and, in cases in which they do not, the coefficients are not statistically significant). Results in Belgium, Cyprus, the Czech Republic, Poland and Spain show similar patterns whenever the coefficients are significant, but statistical significance is less widespread than in the first group of countries above. Estimates for Bulgaria, Slovakia, and the UK show no significant coefficient related to the key parameters of interest; however, the estimates for Slovakia and the UK are consistent with limited commitment when considering the signs of coefficients. Finally, France and Greece present mixed cases because some significant coefficients support limited commitment whereas another significant coefficient contradicts it.

%%% Commitment test
\subsection{Commitment test}

Table \ref{table::test} reports the $p$-values for the various hypothesis tests in \eqref{Eq::Test_final}. \citet{Theloudis2025} show that if one rejects ${\cal H}_{0}^\text{NC}$ (the default null for no commitment), then they would generally also reject ${\cal H}_{0}^\text{FC}$ (the null for full commitment), because full commitment is nested within no commitment in terms of the variables that affect the Pareto weight. This `nesting' of effects is less obvious as soon as one considers the new hypotheses with sign restrictions, therefore each hypothesis should be assessed one-by-one.

\smallskip\smallskip

\begin{table}[htp!]  
\begin{center}
\caption{Commitment test $p$-values}\label{table::test}
\begin{tabular}{L{2.5cm}C{2cm}C{2cm}C{2cm}C{2cm}}
\toprule
           &                          &                          & \multicolumn{2}{c}{new hypotheses}\\
\cmidrule{4-5} 
           & ${\cal H}_{0}^\text{FC}$ & ${\cal H}_{0}^\text{NC}$ & ${\cal H}_{0}^\text{NC-new}$             & ${\cal H}_{0}^\text{LC}$   \\ 
\midrule
Austria    & $<0.001$                 & $<0.001$                 & $<0.001$                                 & $0.686$                    \\
Belgium    & $0.007$                  & $0.024$                  & $0.004$                                  & $0.455$                    \\
Bulgaria   & $0.124$                  & $0.143$                  & $0.216$                                  & $0.344$                    \\
Cyprus     & $0.220$                  & $0.255$                  & $0.354$                                  & $0.826$                    \\
Czech Rep. & $<0.001$                 & $0.075$                  & $0.122$                                  & $0.645$                    \\
France     & $<0.001$                 & $<0.001$                 & $0.001$                                  & $0.147$                    \\
Greece     & $0.002$                  & $0.001$                  & $0.001$                                  & $0.087$                    \\
Hungary    & $0.061$                  & $0.066$                  & $0.110$                                  & $0.989$                    \\
Italy      & $<0.001$                 & $<0.001$                 & $<0.001$                                 & $0.284$                    \\
Luxembourg & $0.009$                  & $0.003$                  & $0.005$                                  & $0.673$                    \\
Poland     & $0.023$                  & $0.010$                  & $0.020$                                  & $0.826$                    \\
Romania    & $<0.001$                 & $0.002$                  & $0.004$                                  & $0.402$                    \\
Slovakia   & $0.157$                  & $0.444$                  & $0.264$                                  & $0.590$                    \\
Spain      & $0.066$                  & $0.031$                  & $0.053$                                  & $0.418$                    \\
UK         & $0.498$                  & $0.381$                  & $0.492$                                  & $0.486$                   \\ 
\bottomrule
\end{tabular}
\begin{minipage}{0.82\textwidth}
\footnotesize
\textit{Notes}: The null hypotheses are formulated in \eqref{Eq::Test_final}. The empirical implementation is based on results from table \ref{table::main}, with details reported in Appendix Tables \ref{table::detailsAT} to \ref{table::detailsUK}.
\end{minipage}
\end{center}
\end{table}

In 9 countries, Austria, Belgium, Czech Republic, France, Greece, Italy, Luxembourg, Poland, and Romania, we reject full commitment strongly. In 2 countries, Hungary and Spain, the $p$-value of this test is just above the $5\%$ level, so we reject full commitment only at the $10\%$. In 4 countries, Bulgaria, Cyprus, Slovakia, and the UK, we cannot reject full commitment; three of them (all except Cyprus) lack statistical significance in the coefficients of table \ref{table::main}, which explains why we cannot reject full commitment there.

In all 11 countries for which we reject full commitment, including Hungary and Spain, we also reject the conventional null for no commitment (${\cal H}_{0}^\text{NC}$). Spain's $p$-value in this case drops below the $5\%$ level but Czech Republic's rises to $0.075$. In the other 4 countries, we cannot reject the conventional null for no commitment.

Assessing no commitment through the lens of the new hypothesis ${\cal H}_{0}^\text{NC-new}$, the 4 countries for which we did not reject the conventional null (Bulgaria, Cyprus, Slovakia, and the UK) also fail to reject the new null. In addition, Spain's $p$-value increases to $0.053$, while Czech Republic's and Hungary's rise to just above $10\%$. The new hypothesis involves the sign of $\beta_{j[\omega_{-jt}]}$, which is not part of the testable restrictions in the conventional ${\cal H}_{0}^\text{NC}$. This coefficient often lacks statistical significance, which makes it harder to reject the new null in our data. 

Moving on to ${\cal H}_{0}^\text{LC}$, we cannot reject limited commitment in any country in the sample, except for Greece for which the $p$-value is just under the $10\%$ level. However, the coefficients on own past shocks in Greece \emph{are} consistent with limited commitment, while the coefficient on the initial distribution factor is not; this generates a tension between rejecting and not rejecting limited commitment, hence this marginal rejection. France is another case in which we previously reported contradictory coefficients; in spite of this, we do not reject limited commitment here. In most other countries, the failure to reject ${\cal H}_{0}^\text{LC}$ does not reflect lack of statistical power; as many estimates in table \ref{table::main} are significant \emph{and} have the sign that limited commitment postulates, household behavior is in fact consistent with limited commitment, which in turn explains why we do not reject ${\cal H}_{0}^\text{LC}$. The only countries where we arguably \emph{do} lack statistical power are Bulgaria, Cyprus, Slovakia, and the UK; but while Slovakia and the UK show no significant effects of wage shocks and distribution factors (hence our failure to reject ${\cal H}_{0}^\text{FC}$ and ${\cal H}_{0}^\text{NC}$ in the first place), their coefficients have the sign postulated by limited commitment. This is further corroborated by the elasticities of the Pareto weight with respect to wages in Section \ref{SubSection::Results_Pareto}.

In summary, we reject full commitment and the conventional null for no commitment in all but 4 countries. These results are in line with \citet{Mazzocco2007}, who reject full commitment in the US, \citet{Lise2019} and \citet{Toriyabe2025}, who also reject it in Japan, and \citet{Theloudis2025}, who reject no commitment in the US. 

Strengthening the test for no commitment by imposing an additional sign restriction still leads to rejection of this regime in the vast majority of countries. Crucially, ours is the first test with power against limited commitment; even so, we do not reject limited commitment in any country, except for one marginal case.

%%% Structural elasticities and the size of bargaining effects
\subsection{The size of bargaining effects}\label{SubSection::Results_Pareto}

The reduced-form labor supply coefficients and the commitment test in the previous sections do not readily quantify the economically relevant bargaining effect of wages, nor do they permit direct cross-country comparisons, as the coefficients depend on observables and past work hours -- see system \eqref{Eq::betas} below. By bargaining effect we refer to the structural elasticities of the Pareto weight with respect to wages, namely $e_{[\mu_j,w_{k}]}$, for spouse $j,k=1,2$, which measure the extent to which the Pareto weight responds to own and cross wage shocks. These elasticities are identified through restrictions across the equations that link the reduced-form coefficients in \eqref{Eq::EstimableEquation} with their underlying structure from \eqref{Eq::FOCs} and \eqref{Eq::Approx_mu}, namely
\begin{align}\label{Eq::betas}
\begin{split}
    \beta_{j[\omega_{jt}]}      =&  \mathbin{\color{white}{-}} (\alpha_j^{-1} + \kappa_t s_{jt-1} h^{-1}_{jt-1})^{-1} \times (1 -\kappa_t s_{jt-1} - e_{[\mu_j,w_{j}]}) \\
    \beta_{j[\omega_{-jt}]}     =& - (\alpha_j^{-1} + \kappa_t s_{jt-1} h^{-1}_{jt-1})^{-1} \times   e_{[\mu_j,w_{-j}]}, \\
    \beta_{j[\omega_{jt-1}]}    =& - (\alpha_j^{-1} + \kappa_t s_{jt-1} h^{-1}_{jt-1})^{-1} \times   e_{[\mu_j,\mu_{jL}]} \times e_{[\mu_j,w_j]}, \\
    \beta_{j[\omega_{-jt-1}]}   =& - (\alpha_j^{-1} + \kappa_t s_{jt-1} h^{-1}_{jt-1})^{-1} \times   e_{[\mu_j,\mu_{jL}]} \times e_{[\mu_j,w_{-j}]}, \\
    \beta_{j[\Delta y_{-jt}]}   =& - (\alpha_j^{-1} + \kappa_t s_{jt-1} h^{-1}_{jt-1})^{-1} \times \kappa_t s_{-jt-1},
\end{split}
\end{align}
where $\beta_{j[\omega_{k\tau}]}$, $j,k=1,2$, are the coefficients on current ($\tau=t$) and past ($\tau=t-1$) wage shocks in \eqref{Eq::EstimableEquation} and $\beta_{j[\Delta y_{-jt}]}$ is the coefficient on the growth rate of partner's earnings $\Delta \log y_{-jt}$. The last coefficient identifies $\alpha_{j}$, the Frisch elasticity of labor supply, since $s_{k}$ and $h_{j}$ (earnings share and hours) are observed in the data, while $\kappa_t$ (tax progressivity parameter) can be obtained externally \citep[see, e.g.,][]{Heathcote2014}. The remaining coefficients then identify $e_{[\mu_j,w_{k}]}$ \emph{and} the elasticity of the Pareto weight with respect to its past value, $e_{[\mu_j,\mu_{jL}]}$. We estimate this system of equations using nonlinear GMM.\footnote{See Appendix \ref{Appendix::Theory} for details on the structure underlying the reduced-form coefficients $\beta_j$, and \citet{Theloudis2025} for the econometric details of this GMM estimation. We set $\kappa_t=0.185$ \citep[as in][for the US]{Heathcote2014}, though our results are reasonably robust to this choice.}

The restrictions on the reduced-form coefficients implied by the commitment test in \eqref{Eq::Test_final} map directly into restrictions on the structural elasticities of the Pareto weight. Under full commitment, $e_{[\mu_j,w_k]}=e_{[\mu_j,\mu_{jL}]}=0$ (the Pareto weight is constant over time). Under no commitment, $e_{[\mu_j,w_j]}>0$ and $e_{[\mu_j,w_{-j}]}<0$ (favorable own shocks empower oneself and weaken the partner) but $e_{[\mu_j,\mu_{jL}]}=0$ (history is irrelevant). In limited commitment, $e_{[\mu_j,w_j]}>0$ and $e_{[\mu_j,w_{-j}]}<0$, and $e_{[\mu_j,\mu_{jL}]}>0$ (the Pareto weight is persistent / has memory). 

\begin{figure}[t!]
    \centering
    \caption{Elasticities of Pareto weights w.r.t. own past shock, $e_{[\mu_j,\mu_{jL}]} \times e_{[\mu_j,w_j]}$}\label{fig::elasticities}
    \includegraphics[width=0.9\linewidth]{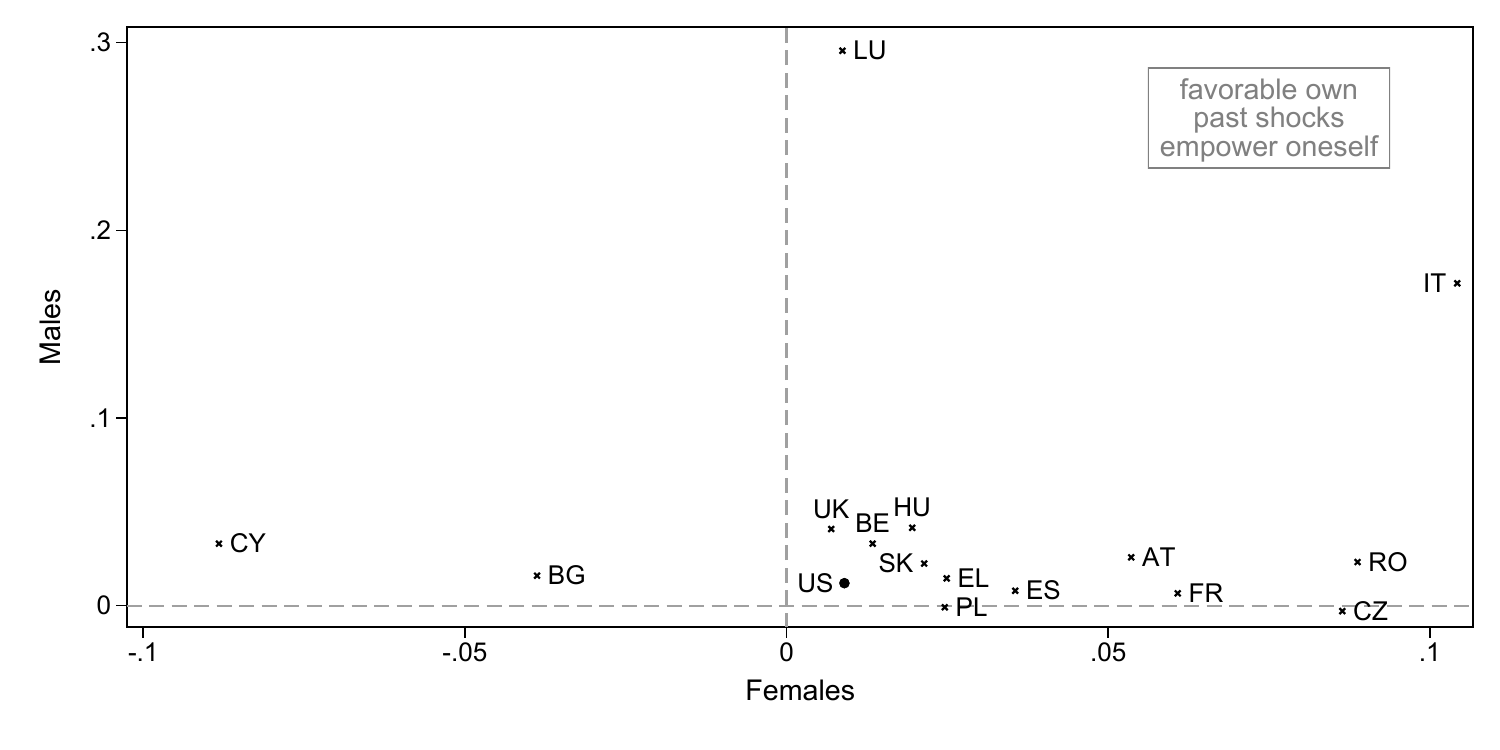}
    \begin{minipage}{0.9\textwidth}
    \footnotesize
    \textit{Notes}: The elasticities are estimated via nonlinear GMM through the system of equations \eqref{Eq::betas}. Elasticities for the US are taken from Table 2 in \citet{Theloudis2025}.
    \end{minipage}
\end{figure}

Figure \ref{fig::elasticities} plots the elasticities of the male and female Pareto weight with respect to own \emph{past} wages, namely $e_{[\mu_j,\mu_{jL}]} \times e_{[\mu_j,w_j]}$, for $j=1,2$. These elasticities speak directly to bargaining through limited commitment and summarize both the role of wage shocks and history for the Pareto weight. Estimates of all elasticities appear in Appendix Table \ref{table::elasticities}. 

Three main points emerge. First, the elasticities lie in the first quadrant for all countries except for Bulgaria, Cyprus, Czech Republic, and Poland. This implies that $e_{[\mu_1,w_1]}>0$ and $e_{[\mu_2,w_2]}>0$ (favorable own shocks empower both men and women) and $e_{[\mu_j,\mu_{jL}]}>0$ (the Pareto weight is persistent), precisely as limited commitment postulates. This applies also to France and Greece (despite the contradicting reduced-form coefficients earlier) and Slovakia and the UK (despite the lack of statistical power to reject any commitment regime previously). Second, the Czech Republic and Poland lie on the border between the first and fourth quadrants; in this case, $e_{[\mu_1,w_1]}=0$ but $e_{[\mu_2,w_2]}>0$ (favorable own shocks empower the women), while $e_{[\mu_j,\mu_{jL}]}>0$ (the Pareto weight is persistent). Third, Bulgaria and Cyprus lie in the second quadrant; this implies $e_{[\mu_1,w_1]}>0$ but $e_{[\mu_2,w_2]}<0$, i.e., favorable own shocks empower men but weaken women, which is not consistent with bargaining. For both countries, we previously lacked statistical power to reject any commitment regime. 

For most countries in the first quadrant, the elasticities for men and women lie approximately between $0.00$-$0.10$. A $10\%$ wage change in the past period thus shifts current hours by up to $1\%$ exclusively through revisions in the Pareto weight. This bargaining effect of wages on labor supply is evidently smaller than the substitution effect (Frisch elasticities are typically around $0.50$ or larger) or the income effect (Marshallian elasticities are typically close to $0.00$ so the income effect is as large as the substitution effect). Italy and Luxembourg are clearly exceptions, since the bargaining effect there is as large as $0.30$.

Appendix Table \ref{table::elasticities} shows that, while the Pareto weight elasticity with respect to own wages is statistically significant (the elasticities we plot above), the elasticity with respect to the partner's wage is not. However, when these cross-elasticities \emph{are} significant, then they have the sign that bargaining predicts.

%%% Extensions
\subsection{Extensions}

\paragraph{Robustness checks} We conduct three robustness checks. First, we restrict the sample to legally married couples, versus married and cohabiting in the baseline, as the two types may have access to different commitment technologies. Second, we limit the sample to couples with children, as childless couples may feature less commitment. Third, we restrict the sample to couples aged 30 to 60, a period in life when most people participate in the labor market. Appendix Tables \ref{table::test_married}, \ref{table::test_wchildren}, and \ref{table::test_3060} present the results. Across these checks, the results remain qualitatively unchanged. We find slightly less evidence against full commitment among married couples or couples with children, but we generally reject full and no commitment in the majority of countries while we cannot reject limited commitment.\footnote{We have also estimated the labor supply equations using alternative country-specific deflators for the monetary amounts, rather than a common deflator (CPI) across countries. Our results are unchanged.} 

\paragraph{Older wage shocks} The use of older wage shocks from period $t-2$ strengthens all parts of the commitment test by introducing additional equality and sign restrictions on the coefficients on those shocks \citep{Theloudis2025}. Doing so, however, requires observing households for an additional year. Given the rotating panel design of the EU-SILC, only a fraction of households are observed for four consecutive years, versus three in our baseline. Specifically, the number of observations entering the estimating equations drops from 65,716 to 26,032 in this case (a 60\% reduction). We conduct the test for commitment including the restrictions from shocks at $t-2$ and present the results in Appendix Table \ref{table::test_historical}. Our main qualitative conclusions remain intact. We reject full and no commitment, but cannot reject limited commitment, in Austria, Belgium, Czech Republic, France, Greece, Italy, and Luxembourg. In Romania we only reject full commitment, while in Cyprus, interestingly, we reject all commitment alternatives. We lack the statistical power to reject any hypothesis in the remaining countries because this specification limits the number of observations and increases the flexibility of the estimating equations.

\paragraph{Bias from omitting the consumption control} Many household surveys do not include information on consumption, which raises concerns about how omitting this control variable may bias our test for commitment. To address this issue, one may use the omitted-variable bias formula in the context of the estimating equation \eqref{Eq::EstimableEquation}, sign the bias in the various coefficients from omitting consumption, and recast the commitment test in terms of coefficients whose sign may flip because of the bias. The idea is that if the coefficients have the sign that limited (no) commitment postulates in spite of the bias to the opposite, this serves as evidence for limited (no) commitment.\footnote{In the context of the EU-SILC data, the bias from omitting consumption is at best very small. This is evident from $b_{j[q_t]}$, the coefficient on consumption in \eqref{Eq::EstimableEquation}, being statistically insignificant across most countries in the sample -- see Appendix Tables \ref{table::detailsAT} to \ref{table::detailsUK}.} This approach thus enables the econometrician to bypass the lack of consumption data and test for commitment. In many cases, this argument can also be generalized to settings in which wealth or leisure data are unavailable.

\paragraph{Heterogeneity} There are many reasons why, even within the same country, households may differ in the extent of commitment and risk sharing between spouses. An advantage of our test is that, in principle, one may implement it on a household by household basis. To see this, inspect the main estimating equation \eqref{Eq::EstimableEquation}. With sufficiently long time series, one may estimate household-specific coefficients $\beta_{j[\omega_{k\tau}]}$ on the wage shocks, $j,k=1,2$, and test the various commitment hypotheses in \eqref{Eq::Test_final} separately for each household.\footnote{This approach does not allow estimation of the coefficient $\beta_{j[\theta_{j}]}$ on the time-invariant initial distribution factor, but this parameter is not crucial for testing the commitment hypotheses.} This would allow us to count how many households operate in each commitment regime and elicit the characteristics of the respective subpopulations. In practice, this is not feasible in our data (the EU-SILC surveys households for up to four years) but it is a promising idea that we leave for future research.

%%%%%%%%%%%%%%%%%%%%%%%%%%%%%%%%%%%%%%%%
%%% Conclusions
%%%%%%%%%%%%%%%%%%%%%%%%%%%%%%%%%%%%%%%%

\section{Conclusions}\label{Section::Conclusion}

This paper studies the extent of commitment in the household across 15 European countries. We propose new tests that distinguish between three commitment types (full, no, and limited commitment) based on the dynamic effects of wage shocks on household labor supply. The distinctive feature of our approach is its ability to formally reject limited commitment, in addition to the other two types, exploiting sign restrictions from theory. Empirically, we find that the elasticity of the Pareto weight with respect to favorable past wages is generally positive, consistent with bargaining under limited
commitment. Formally, we reject full and no commitment in all but 4 countries, but fail to reject limited commitment.

Understanding the dynamics of household behavior is crucial to understanding the effects of policies targeted to the household. Our findings suggest that shocks to the economic environment may have long term effects on the allocation of household resources in several European countries. Consequently, interventions targeting individual income can have broader (intended and unintended) consequences over time. For example, wage subsidies, tax reforms, or policies aimed at strengthening gender parity (e.g., closing gender wage gaps) can generally change intra-household bargaining power in a persistent way. Relatedly, programs that mitigate negative income shocks (such as unemployment insurance) might also serve to smoothen changes in intra-household allocations, thus reducing the incidence of internal renegotiation. The design of family policies and divorce laws must thus consider that intra-household bargaining power is sensitive to economic shocks and evolves over time.

%%%%%%%%%%%%%%%%%%%%%%%%%%%%%%%%%%%%%%%%
%%% References
%%%%%%%%%%%%%%%%%%%%%%%%%%%%%%%%%%%%%%%%

\begin{singlespace}
\bibliographystyle{chicago}
\bibliography{Bibliography_Commitment_Europe}

\begin{thebibliography}{}

\bibitem[\protect\citeauthoryear{Altonji, Smith~Jr, and Vidangos}{Altonji
  et~al.}{2013}]{Altonji2013}
Altonji, J., A.~Smith~Jr, and I.~Vidangos (2013).
\newblock Modelling earnings dynamics.
\newblock {\em Econometrica\/}~{\em 81\/}(4), 1395--1454.

\bibitem[\protect\citeauthoryear{Andrews and Soares}{Andrews and
  Soares}{2010}]{Andrews2010}
Andrews, D.~W. and G.~Soares (2010).
\newblock Inference for parameters defined by moment inequalities using
  generalized moment selection.
\newblock {\em Econometrica\/}~{\em 78\/}(1), 119--157.

\bibitem[\protect\citeauthoryear{Apps and Jones}{Apps and
  Jones}{1986}]{Apps1986}
Apps, P.~F. and G.~Jones (1986).
\newblock Selective taxation of couples.
\newblock {\em Journal of Economics\/}~{\em 5}, 1--15.

\bibitem[\protect\citeauthoryear{Blasutto and Kozlov}{Blasutto and
  Kozlov}{2025}]{Blasutto2025}
Blasutto, F. and E.~Kozlov (2025).
\newblock {(Changing) Marriage and cohabitation patterns in the US: Do divorce
  laws matter?}
\newblock Working Paper 2025-02, ECARES.

\bibitem[\protect\citeauthoryear{Blau and Goodstein}{Blau and
  Goodstein}{2016}]{Blau2016}
Blau, D. and R.~Goodstein (2016).
\newblock Commitment in the household: Evidence from the effect of inheritances
  on the labor supply of older married couples.
\newblock {\em Labour Economics\/}~{\em 42}, 123--137.

\bibitem[\protect\citeauthoryear{Blundell, Pistaferri, and
  Saporta-Eksten}{Blundell et~al.}{2016}]{Blundell2016}
Blundell, R., L.~Pistaferri, and I.~Saporta-Eksten (2016).
\newblock Consumption inequality and family labor supply.
\newblock {\em {American Economic Review}\/}~{\em 106\/}(2), 387--435.

\bibitem[\protect\citeauthoryear{Bourguignon, Browning, and
  Chiappori}{Bourguignon et~al.}{2009}]{Bourguignon2009}
Bourguignon, F., M.~Browning, and P.-A. Chiappori (2009).
\newblock Efficient intra-household allocations and distribution factors:
  Implications and identification.
\newblock {\em Review of Economic Studies\/}~{\em 76\/}(2), 503--528.

\bibitem[\protect\citeauthoryear{Bredemeier, Gravert, and Juessen}{Bredemeier
  et~al.}{2023}]{Bredemeier2023}
Bredemeier, C., J.~Gravert, and F.~Juessen (2023).
\newblock Accounting for limited commitment between spouses when estimating
  labor-supply elasticities.
\newblock {\em Review of Economic Dynamics\/}~{\em 51}, 547--578.

\bibitem[\protect\citeauthoryear{Bronson}{Bronson}{2014}]{Bronson2014}
Bronson, M.~A. (2014).
\newblock {Degrees are forever: Marriage, educational investment, and lifecycle
  labor decisions of men and women}.

\bibitem[\protect\citeauthoryear{Browning, Chiappori, and Weiss}{Browning
  et~al.}{2014}]{Browning2014}
Browning, M., P.-A. Chiappori, and Y.~Weiss (2014).
\newblock {\em Economics of the Family}.
\newblock Cambridge University Press.

\bibitem[\protect\citeauthoryear{Cameron and Miller}{Cameron and
  Miller}{2015}]{Cameron2015}
Cameron, A.~C. and D.~L. Miller (2015).
\newblock A practitioner{\textquoteright}s guide to cluster-robust inference.
\newblock {\em Journal of Human Resources\/}~{\em 50\/}(2), 317--372.

\bibitem[\protect\citeauthoryear{Chernozhukov, Hong, and Tamer}{Chernozhukov
  et~al.}{2007}]{Chernozhukov2007}
Chernozhukov, V., H.~Hong, and E.~Tamer (2007).
\newblock Estimation and confidence regions for parameter sets in econometric
  models.
\newblock {\em Econometrica\/}~{\em 75\/}(5), 1243--1284.

\bibitem[\protect\citeauthoryear{Chiappori}{Chiappori}{1988}]{Chiappori1988}
Chiappori, P.-A. (1988).
\newblock Rational household labor supply.
\newblock {\em Econometrica\/}~{\em 56\/}(1), 63--90.

\bibitem[\protect\citeauthoryear{Chiappori}{Chiappori}{1992}]{Chiappori1992}
Chiappori, P.-A. (1992).
\newblock Collective labor supply and welfare.
\newblock {\em Journal of Political Economy\/}~{\em 100\/}(3), 437--467.

\bibitem[\protect\citeauthoryear{Chiappori, Dias, and Meghir}{Chiappori
  et~al.}{2018}]{Chiappori2018}
Chiappori, P.-A., M.~C. Dias, and C.~Meghir (2018).
\newblock {The marriage market, labor supply, and education choice}.
\newblock {\em Journal of Political Economy\/}~{\em 126\/}(S1), 26--72.

\bibitem[\protect\citeauthoryear{Chiappori, Fortin, and Lacroix}{Chiappori
  et~al.}{2002}]{Chiappori2002}
Chiappori, P.-A., B.~Fortin, and G.~Lacroix (2002).
\newblock Marriage market, divorce legislation, and household labor supply.
\newblock {\em Journal of Political Economy\/}~{\em 110\/}(1), 37--72.

\bibitem[\protect\citeauthoryear{Chiappori, Giménez-Nadal, Molina, and
  Velilla}{Chiappori et~al.}{2022}]{Chiappori2022review}
Chiappori, P.-A., J.~I. Giménez-Nadal, J.~A. Molina, and J.~Velilla (2022).
\newblock Household labor supply: Collective evidence in developed countries.
\newblock In K.~F. Zimmermann (Ed.), {\em Handbook of Labor, Human Resources
  and Population Economics}. Springer.

\bibitem[\protect\citeauthoryear{Chiappori and Mazzocco}{Chiappori and
  Mazzocco}{2017}]{Chiappori2017}
Chiappori, P.-A. and M.~Mazzocco (2017).
\newblock Static and intertemporal household decisions.
\newblock {\em Journal of Economic Literature\/}~{\em 55\/}(3), 985--1045.

\bibitem[\protect\citeauthoryear{Cosaert, Theloudis, and Verheyden}{Cosaert
  et~al.}{2023}]{Cosaert2023}
Cosaert, S., A.~Theloudis, and B.~Verheyden (2023).
\newblock Togetherness in the household.
\newblock {\em American Economic Journal: Microeconomics\/}~{\em 15\/}(1),
  529--579.

\bibitem[\protect\citeauthoryear{De~Nardi, French, Jones, and McGee}{De~Nardi
  et~al.}{2025}]{DeNardiFrenchJonesMcGee2025_WhySave}
De~Nardi, M., E.~French, J.~B. Jones, and R.~McGee (2025).
\newblock {Why do couples and singles save during retirement? Household
  heterogeneity and its aggregate implications}.
\newblock {\em Journal of Political Economy\/}~{\em 133\/}(3), 750--792.

\bibitem[\protect\citeauthoryear{Duflo}{Duflo}{2003}]{Duflo2003}
Duflo, E. (2003).
\newblock {Grandmothers and granddaughters: Old-age pensions and intrahousehold
  allocation in South Africa}.
\newblock {\em World Bank Economic Review\/}~{\em 17\/}(1), 1--25.

\bibitem[\protect\citeauthoryear{Fernández and Wong}{Fernández and
  Wong}{2017}]{Fernandez2017}
Fernández, R. and J.~C. Wong (2017).
\newblock {Free to leave? A welfare analysis of divorce regimes}.
\newblock {\em American Economic Journal: Macroeconomics\/}~{\em 9\/}(3),
  72--115.

\bibitem[\protect\citeauthoryear{Foerster}{Foerster}{2025}]{Foerster2024}
Foerster, H. (2025).
\newblock Untying the knot: How child support and alimony affect couples’
  dynamic decisions and welfare.
\newblock {\em Review of Economic Studies\/}~{\em 92\/}(5), 3029--3066.

\bibitem[\protect\citeauthoryear{Gale and Shapley}{Gale and
  Shapley}{1962}]{Gale1962}
Gale, D. and L.~S. Shapley (1962).
\newblock College admissions and the stability of marriage.
\newblock {\em American Mathematical Monthly\/}~{\em 69\/}(1), 9--15.

\bibitem[\protect\citeauthoryear{Gouss{\'{e}}, Jacquemet, and
  Robin}{Gouss{\'{e}} et~al.}{2017}]{Gousse2017}
Gouss{\'{e}}, M., N.~Jacquemet, and J.~Robin (2017).
\newblock Marriage, labor supply, and home production.
\newblock {\em Econometrica\/}~{\em 85\/}(6), 1873--1919.

\bibitem[\protect\citeauthoryear{Gr{\"o}mping}{Gr{\"o}mping}{2010}]{Gromping2010}
Gr{\"o}mping, U. (2010).
\newblock {Inference with linear equality and inequality constraints using R:
  The package ic. infer}.
\newblock {\em Journal of Statistical Software\/}~{\em 33\/}(1), 1--31.

\bibitem[\protect\citeauthoryear{Heathcote, Storesletten, and
  Violante}{Heathcote et~al.}{2014}]{Heathcote2014}
Heathcote, J., K.~Storesletten, and G.~L. Violante (2014).
\newblock {Consumption and labor supply with partial insurance: An analytical
  framework}.
\newblock {\em American Economic Review\/}~{\em 104\/}(7), 2075--2126.

\bibitem[\protect\citeauthoryear{Jayachandran and Voena}{Jayachandran and
  Voena}{2025}]{JayachandranVoena2025_WomensPower}
Jayachandran, S. and A.~Voena (2025).
\newblock Women's power in the household.
\newblock Working Paper 34605, National Bureau of Economic Research.

\bibitem[\protect\citeauthoryear{Ligon, Thomas, and Worrall}{Ligon
  et~al.}{2002}]{LigonThomasWorrall2002}
Ligon, E., J.~P. Thomas, and T.~Worrall (2002).
\newblock Informal insurance arrangements with limited commitment: Theory and
  evidence from village economies.
\newblock {\em Review of Economic Studies\/}~{\em 69\/}(1), 209--244.

\bibitem[\protect\citeauthoryear{Lise and Yamada}{Lise and
  Yamada}{2019}]{Lise2019}
Lise, J. and K.~Yamada (2019).
\newblock {Household sharing and commitment: Evidence from panel data on
  individual expenditures and time use}.
\newblock {\em Review of Economic Studies\/}~{\em 86\/}(5), 2184--2219.

\bibitem[\protect\citeauthoryear{Low, Meghir, Pistaferri, and Voena}{Low
  et~al.}{2018}]{Low2018}
Low, H., C.~Meghir, L.~Pistaferri, and A.~Voena (2018).
\newblock Marriage, labor supply and the dynamics of the social safety net.
\newblock Working Paper 24356, National Bureau of Economic Research.

\bibitem[\protect\citeauthoryear{Lundberg, Pollak, and Wales}{Lundberg
  et~al.}{1997}]{Lundberg1997}
Lundberg, S., R.~Pollak, and T.~Wales (1997).
\newblock {Do husbands and wives pool their resources? Evidence from the United
  Kingdom child benefit}.
\newblock {\em Journal of Human Resources\/}~{\em 32\/}(3), 463--480.

\bibitem[\protect\citeauthoryear{Lundberg and Pollak}{Lundberg and
  Pollak}{1993}]{LundbergPollak1993_Spheres}
Lundberg, S. and R.~A. Pollak (1993).
\newblock Separate spheres bargaining and the marriage market.
\newblock {\em Journal of political Economy\/}~{\em 101\/}(6), 988--1010.

\bibitem[\protect\citeauthoryear{Lundberg and Pollak}{Lundberg and
  Pollak}{2003}]{LundbergPollak2003}
Lundberg, S. and R.~A. Pollak (2003).
\newblock Efficiency in marriage.
\newblock {\em Review of Economics of the Household\/}~{\em 1\/}(3), 153--167.

\bibitem[\protect\citeauthoryear{Manser and Brown}{Manser and
  Brown}{1980}]{Manser1980}
Manser, M. and M.~Brown (1980).
\newblock Marriage and household decision-making: A bargaining analysis.
\newblock {\em International Economic Review\/}~{\em 21\/}(1), 31.

\bibitem[\protect\citeauthoryear{Mazzocco}{Mazzocco}{2007}]{Mazzocco2007}
Mazzocco, M. (2007).
\newblock {Household intertemporal behaviour: A collective characterization and
  a test of commitment}.
\newblock {\em Review of Economic Studies\/}~{\em 74\/}(3), 857--895.

\bibitem[\protect\citeauthoryear{McElroy and Horney}{McElroy and
  Horney}{1981}]{McElroy1981}
McElroy, M.~B. and M.~J. Horney (1981).
\newblock Nash-bargained household decisions: Toward a generalization of the
  theory of demand.
\newblock {\em International Economic Review\/}~{\em 22\/}(2), 333.

\bibitem[\protect\citeauthoryear{Penglase and S\"{o}zbir}{Penglase and
  S\"{o}zbir}{2026}]{PenglaseSozbir2026}
Penglase, J. and O.~F. S\"{o}zbir (2026).
\newblock {Are households Pareto efficient? A test based on multiple job
  holding}.
\newblock {\em Labour Economics\/}~{\em 98}, 102841.

\bibitem[\protect\citeauthoryear{Reynoso}{Reynoso}{2024}]{Reynoso2024}
Reynoso, A. (2024).
\newblock The impact of divorce laws on the equilibrium in the marriage market.
\newblock {\em Journal of Political Economy\/}~{\em 132\/}(12), 4155--4204.

\bibitem[\protect\citeauthoryear{Scholz, Seshadri, and Khitatrakun}{Scholz
  et~al.}{2006}]{Scholz2006}
Scholz, J.~K., A.~Seshadri, and S.~Khitatrakun (2006).
\newblock {Are Americans saving “optimally” for retirement?}
\newblock {\em Journal of Political Economy\/}~{\em 114\/}(4), 607--643.

\bibitem[\protect\citeauthoryear{Shapley and Shubik}{Shapley and
  Shubik}{1971}]{Shapley1971}
Shapley, L.~S. and M.~Shubik (1971).
\newblock {The assignment game I: The core}.
\newblock {\em International Journal of Game Theory\/}~{\em 1\/}(1), 111--130.

\bibitem[\protect\citeauthoryear{Theloudis, Velilla, Chiappori, Gimenez-Nadal,
  and Molina}{Theloudis et~al.}{2025}]{Theloudis2025}
Theloudis, A., J.~Velilla, P.-A. Chiappori, J.~I. Gimenez-Nadal, and J.~A.
  Molina (2025).
\newblock Commitment and the dynamics of household labor supply.
\newblock {\em The Economic Journal\/}~{\em 135\/}(665), 354--386.

\bibitem[\protect\citeauthoryear{Toriyabe}{Toriyabe}{2025}]{Toriyabe2025}
Toriyabe, T. (2025).
\newblock {Empowerment effects and intertemporal commitment of married couples:
  Evidence from Japanese pension reform}.
\newblock {\em Japanese Economic Review\/}.

\bibitem[\protect\citeauthoryear{Voena}{Voena}{2015}]{Voena2015}
Voena, A. (2015).
\newblock {Yours, mine, and ours: Do divorce laws affect the intertemporal
  behavior of married couples?}
\newblock {\em American Economic Review\/}~{\em 105\/}(8), 2295--2332.

\bibitem[\protect\citeauthoryear{Wolak}{Wolak}{1987}]{Wolak1987}
Wolak, F.~A. (1987).
\newblock An exact test for multiple inequality and equality constraints in the
  linear regression model.
\newblock {\em Journal of the American Statistical Association\/}~{\em
  82\/}(399), 782--793.

\bibitem[\protect\citeauthoryear{Wolak}{Wolak}{1989}]{Wolak1989}
Wolak, F.~A. (1989).
\newblock Testing inequality constraints in linear econometric models.
\newblock {\em Journal of Econometrics\/}~{\em 41\/}(2), 205--235.

\end{thebibliography}
\end{singlespace}

%%%%%%%%%%%%%%%%%%%%%%%%%%%%%%%%%%%%%%%%
%%% Appendix settings
%%%%%%%%%%%%%%%%%%%%%%%%%%%%%%%%%%%%%%%%

% The appendix command is issued once, prior to all appendices, if any.
\newpage
\let\origappendix\appendix % save the existing appendix command
\renewcommand\appendix{\clearpage\pagenumbering{arabic}\origappendix}
\appendix
\renewcommand\appendixtocname{Appendix}
\renewcommand\appendixpagename{Appendix}
\appendixpage           % Adds an 'Appendix' title before the first appendix
\addappheadtotoc        % Adds an 'Appendix' title in the table of contents
%   Separate enumeration in the appendix:
\numberwithin{equation}{section} 
\numberwithin{table}{section}   
\numberwithin{figure}{section}
\setcounter{footnote}{0}

%%%%%%%%%%%%%%%%%%%%%%%%%%%%%%%%%%%%%%%%
%%% Appendix A: Theory
%%%%%%%%%%%%%%%%%%%%%%%%%%%%%%%%%%%%%%%%

\section{Theoretical notes and derivations}\label{Appendix::Theory}

%%% Recursive formulation of household problem
\subsection*{Recursive formulation and optimality conditions}

In a previous paper, \citet{Theloudis2025}, we show that the full, limited, and no commitment household programs can be cast into a common recursive formulation, given by
\begin{align*}
    V_t = \max\limits_{\mathbf{C_t}}    & \sum\limits_j \left( \mu_{j(t)} u_j(q_t, l_t, h_{jt}) - \nu_{jt}\tilde{V}_{jt} \right) + \beta \mathds{E}_t V_{t+1} \\
    \text{s.t. }                        & (1+r) a_t + \tau(y_t) = q_t + a_{t+1}, \text{ and} \\
                                        &\text{restrictions on the Pareto weight given by \eqref{Eq::Nesting},}
\end{align*}
where $\nu_{jt}$ is the Lagrange multiplier on spouse $j$'s participation constraint at time $t$ under limited commitment, and $\tilde{V}_{jt}$ is their outside option ($\nu_{jt}=0$ in the other regimes). The optimal labor supply policies $h_{1t}^*$ and $h_{2t}^*$ that solve this problem are functions of the state space, $\mathbf{\Omega}_t = \{w_{1t}, w_{2t}, a_t, \mu_{j(t)}\}$. As a consequence, the restrictions on the Pareto weight in \eqref{Eq::Nesting} are also restrictions on the spouses' optimal choices $h_{1t}^*$ and $h_{2t}^*$. 

We estimate $h_{1t}^*$ and $h_{2t}^*$ leaving preferences and expectations unspecified, following \citet{Blundell2016}. In doing so, we derive the static first-order conditions for individual labor supply, assuming $\tau(y_t) \approx (1-\chi_t) y_t^{1-\kappa_t}$ as in \citet{Heathcote2014}. Utilities can be expressed as $u_j(q_t, l_t, h_{jt}; \boldsymbol{\xi}_{jt}) = \ddot{u}_j(\ddot{q}_t, \ddot{l}_t, \ddot{h}_{jt}),$ with $\boldsymbol{\xi}_{jt}$ being a vector of observed taste shifters, $\ddot{q}_t = q_t e^{-\boldsymbol{\pi}_j^{q\prime} \boldsymbol{\xi}_{jt}}$, $\ddot{l}_t = l_t e^{-\boldsymbol{\pi}_j^{l\prime} \boldsymbol{\xi}_{jt}}$, $\ddot{h}_{jt} = h_{jt} e^{-\boldsymbol{\pi}_j^{h\prime} \boldsymbol{\xi}_{jt}}$, and the $\boldsymbol{\pi}$'s being the loading factors of the taste observables onto choices. The static first-order condition for spouse $j$'s hours is then given by
\begin{equation*}
    -\mu_{j(t)} \dfrac{\partial \ddot{u}_j}{\partial h_j} e^{-\boldsymbol{\pi}_j^{h\prime} \boldsymbol{\xi}_{jt}} = \lambda_t (1-\chi_t)(1-\kappa_t) y_t^{-\kappa_t} w_{jt}.
\end{equation*}
Taking logs and a first difference in time yields $\Delta\log(-\partial \ddot{u}_j/\partial h_j) = \boldsymbol{\pi}_j^{h\prime} \Delta \boldsymbol{\xi}_{jt} + \Delta\log\lambda_t + \Delta\log (1-\chi_t)(1-\kappa_t) - \Delta\kappa_t \log y_t + \Delta\log w_{jt} - \Delta \log \mu_{j(t)}$. We then follow \citet{Blundell2016} and expand the log marginal utility around consumption, leisure, and hours choices in the past (i.e., $t-1$). Putting everything together yields the labor supply function 
\begin{align*}
\Delta\log h_{jt} \approx \alpha_j h_{jt-1}^{-1} \big\{ & \boldsymbol{\pi}_j^{h\prime} \Delta \boldsymbol{\xi}_{jt} + \Delta\log (1-\chi_t)(1-\kappa_t) - \Delta\kappa_t \log y_t + \Delta\log\lambda_t \\
                                                        &- \zeta^q_j q_{t-1} \Delta\log q_t - \zeta^l_j l_{t-1} \Delta\log l_t - \Delta\log\mu_{j(t)} \big\},
\end{align*}
where 
$\alpha_j = \frac{\partial \ddot{u}_j/\partial h_j}{\partial^2 \ddot{u}_j/\partial h_j^2}e^{\boldsymbol{\pi}_j^{h\prime} \boldsymbol{\xi}_{jt}}>0$ is approximately equal to $j$'s Frisch elasticity of labor supply scaled by their work hours, while $\zeta_j^q = \frac{\partial^2 \ddot{u}_j/\partial h_j \partial q}{\partial \ddot{u}_j/\partial h_j}e^{-\boldsymbol{\pi}_j^{q\prime} \boldsymbol{\xi}_{jt}}$ and $\zeta_j^l = \frac{\partial^2 \ddot{u}_j/\partial h_j \partial l}{\partial \ddot{u}_j/\partial h_j}e^{-\boldsymbol{\pi}_j^{l\prime} \boldsymbol{\xi}_{jt}}$ reflect the complementarity between labor supply and consumption/joint leisure, respectively. Assuming the progressivity parameter $\kappa_t$ does not change between proximate time periods, and writing $\Delta\kappa_t \log y_t \approx \kappa_{t-1}(s_{jt-1}(\Delta\log w_{jt} + \Delta\log h_{jt}) + s_{-jt-1}\Delta\log y_{-jt})$, yields the expression for $\Delta\log h_{jt}$ in \eqref{Eq::FOCs}.

%%% Dynamics of Pareto weight
\subsection*{Dynamics of the Pareto weight}

In terms of the set of variables (information set) that matter for the Pareto weight, the limited commitment weight is the most general. In fact, the full commitment weight is strictly nested within the no commitment weight, which in turn is strictly nested within the limited commitment weight \citep{Theloudis2025}. Thus, in what follows we consider the most general (limited commitment) Pareto weight, given by $\mu_{jt} = \mu_{jt-1} + \nu_{jt}(w_{1t},w_{2t},\mathbf{Z}_{t},a_{t},\mu_{jt-1})$ for $t\geq1$, with $\mu_{j0} = \mu_j (\mathbf{\Theta}_{0})$. We can write $\mu_{jt}$ compactly as $\mu_{jt} = \mu_j (w_{1t},w_{2t},\mathbf{Z}_{t},a_{t},\mu_{jt-1})$ for $t\geq1$, for some appropriate function $\mu_j$.

This general Pareto weight has memory, summarized by the past weight $\mu_{jt-1}$, which accounts for the history of the household from $t=0$ until the current date. The weights are unobserved, so we substitute away the past weights recursively until $t=0$, namely
\begin{align*}
    \mu_{jt} 
    &= \mu_j (w_{1t},w_{2t},\mathbf{Z}_{t},a_{t}, \overbrace{\mu_j(w_{1t-1},w_{2t-1},\mathbf{Z}_{t-1},a_{t-1}, \underbrace{\mu_j(w_{1t-2},w_{2t-2},\mathbf{Z}_{t-2},a_{t-2}, \dots, \underbrace{\mu_j(\mathbf{\Theta}_0)}_{\mu_{j0}})}_{\mu_{jt-2}})}^{\mu_{jt-1}})\\
    &= \mu_j (w_{1t},w_{2t},\mathbf{Z}_{t},a_{t}, \overbrace{w_{1t-1},w_{2t-1},\mathbf{Z}_{t-1},a_{t-1},\underbrace{w_{1t-2},w_{2t-2},\mathbf{Z}_{t-2},a_{t-2}, \dots, \mathbf{\Theta}_0}_{\text{enters through $\mu_{jt-2}$}}}^{\text{enters through $\mu_{jt-1}$}}).
\end{align*}

Now, suppose $\ddot{\mu}_{j}$ is a smooth approximation of $\mu_j$, and let for now the Pareto weight depend only on a stochastic distribution factor $z_{jt} \in \mathbf{Z}_t$ that --without loss of generality-- empowers $j$, and the past weight. That is, $\mu_{jt} \approx \ddot{\mu}_{j}(z_{jt}, \mu_{jt-1})$. The Pareto weight can then be approximated as $\Delta\log\mu_{jt} \approx e_{[\mu_j, z_j]} \Delta\log z_{jt} + e_{[\mu_j, \mu_{jL}]} \Delta\log \mu_{jt-1}$, where $e_{[\mu_j, z_j]}>0$ is the elasticity of the Pareto weight with respect to $z_{jt}$, and $e_{[\mu_j, \mu_{jL}]}>0$ is the elasticity with respect to the past weight. Substituting recursively until $t=0$, one arrives at the initial Pareto weight $\Delta\log \mu_{j0}$, which is driven by some initial distribution factor $\theta_{j0} \in \mathbf{\Theta}_0$. Let $\theta_{j0}$ also empower $j$, so that $\Delta\log \mu_{j0} = e_{[\mu_j, \theta_j]} \theta_{j0},$ where $e_{[\mu_j, \theta_j]}>0$ is the elasticity with respect to $\theta_{j0}$. Assembling everything yields an expression of the Pareto weight at date $t$, namely
\begin{equation*}
    \Delta\log\mu_{jt} \approx \sum\limits_{\tau=0}^{t-1} e_{[\mu_j, \mu_{jL}]}^{\tau} e_{[\mu_j, z_j]} \Delta\log z_{jt-\tau} + e_{[\mu_j, \mu_{jL}]}^{t} e_{[\mu_j, \theta_j]} \theta_{j0}.
\end{equation*}
Returning now to the more general formulation $\mu_{jt} = \mu_j (w_{1t},w_{2t},z_{jt},a_{t},\mu_{jt-1})$ that includes wages and wealth as arguments, one may repeat the previous steps to get
\begin{align*}
\Delta\log \mu_{jt} \approx \sum\limits_{\tau=0}^{t-1} e_{[\mu_j, \mu_{jL}]}^\tau \Big( & e_{[\mu_j, w_1]} \Delta\log w_{1t-\tau} + e_{[\mu_j, w_2]} \Delta\log w_{2t-\tau} \\
& + e_{[\mu_j, z_j]} \Delta\log z_{jt-\tau} + e_{[\mu_j, a]} \Delta\log a_{t-\tau} \Big) + e_{[\mu_j, \mu_{jL}]}^t e_{[\mu_j, \theta_j]} \theta_{j0},
\end{align*}
which is the compact counterpart of expression \eqref{Eq::Approx_mu} in the text. 

This approximation corresponds to the limited commitment Pareto weight in which wages, wealth, stochastic distribution factors and history matter. Under no commitment, history is irrelevant, so $e_{[\mu_j, \mu_{jL}]}=0$, and only news at date $t$ matter. Finally, under full commitment, neither history nor current news matter, and the Pareto weight is constant over time. In our first-difference approach, $\Delta\log \mu_{jt} = 0$ even if the initial distribution factor $\theta_{j0}$ shapes the formation of the initial weight at the date of marriage.

%%% Dynamics of household labor supply
\subsection*{Dynamics of household labor supply} 

Putting together expression \eqref{Eq::FOCs}, which is the expression for $\Delta\log h_{jt}$ that stems from the first-order condition, and \eqref{Eq::Approx_mu}, which characterizes the dynamics of the Pareto weight, yields an equation for the dynamics of household labor supply, for $j=1,2$, given by
\begin{align*}
\Delta \log h_{jt} 
    &= \delta_{jt} h_{jt-1}^{-1} \boldsymbol{\pi}_j^{h\prime} \Delta \boldsymbol{\xi}_{jt} + \delta_{jt} h_{jt-1}^{-1} \Delta \log (1-\chi_{t}) - \delta_{jt} \kappa_{t} s_{-jt-1} h_{jt-1}^{-1} \Delta \log y_{-jt} \\
    &+ \delta_{jt} h_{jt-1}^{-1} \Delta \log \lambda_{t} - \delta_{jt} \zeta_{j}^q h_{jt-1}^{-1} q_{t-1} \Delta \log q_{t} - \delta_{jt} \zeta_{j}^l h_{jt-1}^{-1} l_{t-1} \Delta \log l_{t} \\
    &+ \underbrace{\delta_{jt} (1-\kappa_{t} s_{jt-1} - e_{[\mu_j, w_j]})}_{\mathclap{\substack{\beta_{j[\omega_{jt}]}:\\\text{substitution and bargaining}\\\text{effects of own current wage}}}} h_{jt-1}^{-1} \Delta \log w_{jt} \\
    &- \sum\limits_{\tau=1}^{t-1} \underbrace{\delta_{jt} e_{[\mu_j, \mu_{jL}]}^\tau e_{[\mu_j, w_j]}}_{\mathclap{\substack{\beta_{j[\omega_{jt-\tau}]}:\\\text{bargaining effects}\\\text{of own past wages}}}} h_{jt-1}^{-1} \Delta \log w_{jt-\tau} %
    - \sum\limits_{\tau=0}^{t-1} \underbrace{\delta_{jt} e_{[\mu_j, \mu_{jL}]}^\tau e_{[\mu_j, w_{-j}]}}_{\mathclap{\substack{\beta_{j[\omega_{-jt-\tau}]}:\\\text{bargaining effects}\\\text{of partner's wages}}}} h_{jt-1}^{-1} \Delta \log w_{-jt-\tau} \\
    &- \sum\limits_{\tau=0}^{t-1} \underbrace{\delta_{jt} e_{[\mu_j, \mu_{jL}]}^\tau e_{[\mu_j, z_{j}]}}_{\mathclap{\substack{\beta_{j[z_{jt-\tau}]}:\\\text{bargaining effects of}\\\text{stochastic distribution factors}}}} h_{jt-1}^{-1} \Delta \log z_{jt-\tau} %
    - \sum\limits_{\tau=0}^{t-1} \underbrace{\delta_{jt} e_{[\mu_j, \mu_{jL}]}^\tau e_{[\mu_j, a]}}_{\mathclap{\substack{\beta_{j[a_{t-\tau}]}:\\\text{bargaining effects}\\\text{of wealth}}}} h_{jt-1}^{-1} \Delta \log a_{t-\tau} \\
    &- \underbrace{\delta_{jt} (e_{[\mu_j, \mu_{jL}]})^t e_{[\mu_j, \theta_j]}}_{\mathclap{\substack{\beta_{j[\theta_{j}]}:\\\text{bargaining effects of}\\\text{initial distribution factor}}}} h_{jt-1}^{-1} \theta_{j0}, 
\end{align*}
where $\delta_{jt} = (\alpha_{j}^{-1} + \kappa_{t} s_{jt-1} h_{jt-1}^{-1})^{-1}$. Expression \eqref{Eq::Equation} in the text is a compact version of this.

%%% Derivation of estimating equation
\subsection*{Derivation of estimating equation} 

To obtain the estimating equation \eqref{Eq::EstimableEquation}, we further make these modeling choices: we separate the deterministic and stochastic component of wages assuming $\Delta\log w_{jt} = \boldsymbol{\pi}_j^{\prime} \mathbf{x}_{jt}^w + \omega_{jt}$; we focus on wage shocks at $t$ and $t-1$ rather than the entire history of the household; we discard all other stochastic distribution factors $z_{jt} \in \mathbf{Z}_t$; and we approximate the marginal utility of wealth $\lambda_t$ by $\Delta \log \lambda_{t} \approx \xi_1 \Delta\log a_t + \xi_2 \log a_{t-1} + \xi_3 \Delta\log y_t + \xi_4 \log y_{t-1}$. This yields
\begin{align*}
\Delta \log h_{jt} & = b_{j[0]}h_{jt-1}^{-1} + \mathbf{b}_{j[x]}^{\prime}h_{jt-1}^{-1}\mathbf{x}_{jt} - \delta_{jt} \kappa_{t} s_{-jt-1} h_{jt-1}^{-1} \Delta \log y_{-jt} \\
& + \delta_{jt} (\xi_1- e_{[\mu_j, a]}) h_{jt-1}^{-1} \Delta\log a_{t} + \delta_{jt} \xi_2 h_{jt-1}^{-1} \log a_{t-1} - \delta_{jt} e_{[\mu_j, \mu_{jL}]}e_{[\mu_j, a]} h_{jt-1}^{-1} \Delta \log a_{t-1}  \\
& + \delta_{jt} \xi_3 h_{jt-1}^{-1} \Delta\log y_{t} + \delta_{jt} \xi_4 h_{jt-1}^{-1} \log y_{t-1} \\
& - \delta_{jt} \zeta_{j}^q h_{jt-1}^{-1} q_{t-1} \Delta \log q_{t} - \delta_{jt} \zeta_{j}^l h_{jt-1}^{-1} l_{t-1} \Delta \log l_{t} \\
& + \delta_{jt} (1-\kappa_{t} s_{jt-1} - e_{[\mu_j, w_j]}) h_{jt-1}^{-1} \omega_{jt} - \delta_{jt} e_{[\mu_j, w_{-j}]} h_{jt-1}^{-1} \omega_{-jt} \\
& - \delta_{jt} e_{[\mu_j, \mu_{jL}]}e_{[\mu_j, w_j]} h_{jt-1}^{-1} \omega_{jt-1} - \delta_{jt} e_{[\mu_j, \mu_{jL}]}e_{[\mu_j, w_{-j}]}h_{jt-1}^{-1} \omega_{-jt-1} \\
& - \delta_{jt} (e_{[\mu_j, \mu_{jL}]})^t e_{[\mu_j, \theta_j]} h_{jt-1}^{-1} \theta_{j0}, 
\end{align*}
for $j=1,2$, where the first two terms subsume $\Delta \log (1-\chi_{t})$ and the terms that relate to the taste and wage observables. Its reduced-form counterpart \eqref{Eq::EstimableEquation} is given in full by
\begin{align*}
\Delta \log h_{jt} = h_{jt-1}^{-1} & \times \Big\{ b_{j[0]} + \mathbf{b}_{j[x]}^{\prime}\mathbf{x}_{jt} + b_{j[\Delta y_{-jt}]} s_{-jt-1} \Delta \log y_{-jt} \\
    & + b_{j[\Delta a_t]} \Delta \log a_t + b_{j[a_{t-1}]} \log a_{t-1} + b_{j[\Delta a_{t-1}]} \Delta \log a_{t-1} \\
    & + b_{j[\Delta y_t]} \Delta \log y_t + b_{j[y_{t-1}]} \log y_{t-1} \\
    & + b_{j[q_t]} q_{t-1} \Delta \log q_{t} + b_{j[l_t]} l_{-jt-1} \Delta \log l_{-jt} \\
    & + \beta_{j[\omega_{jt}]} \omega_{jt} + \beta_{j[\omega_{-jt}]} \omega_{-jt} + \beta_{j[\omega_{jt-1}]} \omega_{jt-1} + \beta_{j[\omega_{-jt-1}]} \omega_{-jt-1} + \beta_{j[\theta_j]} \theta_{j0} \Big\},
\end{align*}
where we use the partner's leisure ($l_{-jt}$) as proxy for joint leisure ($l_{t}$).

%%%%%%%%%%%%%%%%%%%%%%%%%%%%%%%%%%%%%%%%
%%% Appendix B: Sample, variables and descriptives
%%%%%%%%%%%%%%%%%%%%%%%%%%%%%%%%%%%%%%%%

\section{Sample, variables, and summary statistics}\label{Appendix::Sumstats}

%%% Sample
\subsection*{Sample} 

The raw EU-SILC longitudinal samples include information for Austria, Belgium, Bulgaria, Croatia, Cyprus, the Czech Republic, Denmark, Estonia, Finland, France, Germany, Greece, Hungary, Iceland, Ireland, Italy, Latvia, Lithuania, Luxembourg, Malta, the Netherlands, Norway, Poland, Portugal, Romania, Serbia, Slovakia, Slovenia, Spain, Sweden, Switzerland, and the United Kingdom. However, Germany only appears in the data since 2018, so we cannot include it in the analysis.

Several countries sample individuals, not households. In those cases, only one adult household member aged 16+ is interviewed, resulting in a representative sample of individuals. This is the case for Denmark, Finland, Iceland, the Netherlands, Norway, Slovenia, and Sweden. Key information on the spouse is thus missing, so these countries are omitted from the sample. This leaves Austria, Belgium, Bulgaria, Croatia, Cyprus, the Czech Republic, Estonia, France, Greece, Hungary, Ireland, Italy, Latvia, Lithuania, Luxembourg, Malta, Poland, Portugal, Romania, Serbia, Slovakia, Spain, Switzerland, and the United Kingdom as countries with usable information for the analysis.

From the countries in this sample, we keep only the reference person in each household, and his/her partner, meaning that we retain only two-member households. Then, we keep households in which both partners work. We also require households to be followed for up to three consecutive years, as two periods are needed to define first differences of key variables, and a third consecutive year is required to define past wage shocks. We require valid information on all variables of interest, and we remove outliers using the Blocked Adaptive Computationally Efficient Outlier Nominators algorithm. After these sample adjustments, Croatia, Estonia, Ireland, Latvia, Lithuania, Malta, Portugal, Serbia, and Switzerland end up with very small sample sizes, so we decided to omit them from the analysis. These requirements result in the set of countries shown in Table \ref{AppTable::Sample}.

\begin{table}[h!]  
\begin{center}
\caption{Sample composition}\label{AppTable::Sample}
\begin{tabular}{L{4.25cm} C{2.1cm} C{3cm} C{2.1cm}}
\toprule
 Countries					& Households	& Observations			& Years \\ \midrule
Austria (AT)				& 1,960 			& 5,130 						& 2007-2019 	\\
Belgium (BE) 			& 2,371 			& 6,486 						& 2006-2019	\\
Bulgaria (BG) 			& 1,772 			& 5,272 						& 2007-2019	\\
Cyprus (CY)			& 2,127 			& 5,668 						& 2007-2019	\\
Czechia (CZ)			& 3,721 			& 9,790 						& 2007-2019	\\
France (FR)				& 1,447 			& 3,460	 					& 2007-2019	\\
Greece (EL)				& 1,986 			& 5,081	 					& 2007-2019	\\
Hungary (HU) 		& 1,839 			& 4,560 						& 2007-2019	\\
Italy (IT)					& 4,264 			& 11,043 					& 2007-2019	\\
Luxembourg (LU)	& 1,982 			& 5,757 						& 2007-2019	\\
Poland (PL)				& 5,283 			& 13,671 					& 2005-2019	\\
Romania (RO)			& 2,857 			& 7,796	 					& 2007-2019	\\
Slovakia (SK)			& 2,362 			& 6,877 						& 2006-2019	\\
Spain (ES)				& 3,433 			& 8,578	 					& 2007-2019	\\
United Kingdom (UK)& 2,400 		& 6,244						& 2007-2018	\\ 
\bottomrule
\end{tabular}
\begin{minipage}{0.8\textwidth}
\footnotesize
\textit{Notes}: The samples (EU-SILC longitudinal) are restricted to working couples between 21 and 65 years old, with complete data on key variables, and observed at least for three consecutive years. Observations correspond to households$\times$years.
\end{minipage}
\end{center}
\end{table} 

%%% Variable definitions
\subsection*{Variables} 

We now present details on variable definitions. For each household fulfilling the above sample requirements, formed by spouses $j=1,2$:
\begin{itemize}
    \item Work hours ($h_{jt}$) are defined as the number of hours usually worked per week in all jobs. We define annual hours of work as hours/week $\times$ 52.14 weeks/year.
    \item Earnings ($y_{jt}$) are defined as employee cash or near-cash gross annual earnings.
    \item We define wages as the ratio of annual earnings over hours of work ($w_{jt}=y_{jt}/h_{jt}$).
    \item Family earnings ($y_{t}$) are defined as the sum of annual labor earnings of spouses, $y_{t}=y_{1t}+y_{2t}$. A spouse's share of earnings ($s_{jt}$) is the ratio of individual earnings over family earnings, $s_{jt}=y_{jt}/y_{t}$.
    \item Wealth ($a_{t}$) is not directly observed in the EU-SILC data. We define wealth as the sum of interests, dividends, and profits from capital investment (referring to the amount of interest from assets such as bank accounts, certificates of deposit, bonds, etc., dividends, and profits from capital investment in an unincorporated business, in which the person does not work). We acknowledge that this does not represent wealth, but \emph{returns} to wealth, which we use as a proxy for wealth in the empirical analysis.
    \item Expenditure ($q_{t}$) is defined as the monthly income to make ends meet, and it is obtained from the survey question: \emph{In your opinion, what is the very lowest net monthly income that your household would have to have in order to make ends meet, that is to pay its usual necessary expenses? Please answer in relation to the present circumstances of your household, and what you consider to be usual necessary expenses (to make ends meet).} This variable is intended to represent \emph{usual} monthly expenses. We define household annual expenditures as the annual equivalent of this variable.
    \item Joint leisure ($l_{t}$) is not observed in the EU-SILC, in fact individual leisure is not observed either. We create individual leisure by using the implied time budget, namely total hours minus market hours, housework, and sleep, following \citet{Theloudis2025}. We impute the average hours of housework from the European Working Conditions Survey (EWCS), while we impute the average hours of sleep from the Harmonized European Time Use Survey (HETUS).
\end{itemize}
Summary statistics are presented in Table \ref{AppTable::Sumstatscc}.

\begin{sidewaystable}[h!]
\begin{center}
\caption{Averages of key variables}\label{AppTable::Sumstatscc}
\begin{tabular}{lcccccccc}
\toprule
Variables                  & Austria & Belgium & Bulgaria & Cyprus & Czech Rep. & France & Greece & Hungary \\ \midrule
Age of husband           & 46.96      & 43.70      & 48.48      & 45.26      & 47.32      & 46.40      & 45.43      & 47.12      \\
Age of wife              & 44.61      & 41.75      & 45.40      & 42.29      & 44.86      & 44.55      & 42.05      & 44.46      \\
College ed. husband      & 0.37       & 0.50       & 0.31       & 0.43       & 0.20       & 0.46       & 0.48       & 0.25       \\
College ed. wife         & 0.30       & 0.61       & 0.46       & 0.52       & 0.20       & 0.51       & 0.54       & 0.34       \\
Work hours husband       & 2.19       & 2.09       & 2.13       & 2.14       & 2.22       & 2.16       & 2.08       & 2.12       \\
Work hours wife          & 1.63       & 1.71       & 2.10       & 1.93       & 2.08       & 1.86       & 1.89       & 2.04       \\
Wage husband             & 24.70      & 23.40      & 3.17       & 16.12      & 6.55       & 19.36      & 12.25      & 4.47       \\
Wage wife                & 18.36      & 20.55      & 2.62       & 13.17      & 4.92       & 16.09      & 10.53      & 3.78       \\
\# kids                  & 1.13       & 1.44       & 1.21       & 1.51       & 1.24       & 1.54       & 1.47       & 1.15       \\
Family earnings          & 97.84      & 93.87      & 16.02      & 67.11      & 28.93      & 83.01      & 47.46      & 20.34      \\
Wealth                   & 5.75       & 8.78       & 1.56       & 8.71       & 0.39       & 9.25       & 4.59       & 2.33       \\
Expenditures             & 32.94      & 40.25      & 16.10      & 41.30      & 14.70      & 71.26      & 34.74      & 9.30       \\
N. households            & 1,960       & 2,371       & 1,772       & 2,127       & 3,721       & 1,447       & 1,986       & 1,839       \\
Obs. (households $\times$ year) & 5,130       & 6,486       & 5,272       & 5,668       & 9,790       & 3,460       & 5,081       & 4,560 \\
\bottomrule
\end{tabular}
\begin{minipage}{0.88\textwidth}
\footnotesize
\textit{Notes}: Age is measured in years. College education represents the rate of individuals with college education, or higher. Work hours are defined as hours per year, divided by 1000. All monetary variables are measured in \EUR2019. Wages are measured in \EUR/hour. Family earnings, wealth, and expenditures are measured as \EUR/year, divided by 1000.
\end{minipage}
\end{center}
\end{sidewaystable}

\begin{sidewaystable}[h!]
\begin{center}
\caption*{Table \ref{AppTable::Sumstatscc} (\textit{cont.}): Averages of key variables}
\begin{tabular}{lccccccc}
\toprule
Variables                  & Italy  & Luxembourg & Poland & Romania & Slovakia & Spain & UK\\ \midrule
Age of husband           & 46.91      & 43.54       & 43.36       & 44.83       & 45.53       & 44.76       & 44.44       \\
Age of wife              & 44.36      & 41.44       & 41.51       & 42.31       & 43.20       & 42.86       & 42.37       \\
College ed. husband      & 0.22       & 0.32        & 0.33        & 0.31        & 0.28        & 0.52        & 0.51        \\
College ed. wife         & 0.28       & 0.39        & 0.46        & 0.31        & 0.31        & 0.61        & 0.57        \\
Work hours husband       & 2.08       & 2.22        & 2.18        & 2.14        & 2.17        & 2.14        & 2.23        \\
Work hours wife          & 1.75       & 1.76        & 2.02        & 2.11        & 2.07        & 1.87        & 1.72        \\
Wage husband             & 18.42      & 30.18       & 5.68        & 2.79        & 4.97        & 15.88       & 19.62       \\
Wage wife                & 14.75      & 26.94       & 4.80        & 2.35        & 3.96        & 13.41       & 15.42       \\
\# kids                  & 1.35       & 1.33        & 1.33        & 1.03        & 1.60        & 1.30        & 1.09        \\
Family earnings          & 73.40      & 127.99      & 25.32       & 12.61       & 24.14       & 64.58       & 80.04       \\
Wealth                   & 7.81       & 12.60       & 2.38        & 0.14        & 1.79        & 6.92        & 4.20        \\
Expenditures             & 31.24      & 51.73       & 11.65       & 8.16        & 18.35       & 33.69       & 27.18       \\
N. households            & 4,264       & 1,982        & 5,283        & 2,857        & 2,362        & 3,433        & 2,400        \\
Obs. (households X year) & 11,043      & 5,757        & 13,671       & 7,796        & 6,877        & 8,578        & 6,244    \\
\bottomrule
\end{tabular}
\begin{minipage}{0.79\textwidth}
\footnotesize
\textit{Notes}: Age is measured in years. College education represents the rate of individuals with college education, or higher. Work hours are defined as hours per year, divided by 1000. All monetary variables are measured in \EUR2019. Wages are measured in \EUR/hour. Family earnings, wealth, and expenditures are measured as \EUR/year, divided by 1000.
\end{minipage}
\end{center}
\end{sidewaystable}

\clearpage

%%%%%%%%%%%%%%%%%%%%%%%%%%%%%%%%%%%%%%%%
%%% Appendix C: Results in detail and additional results
%%%%%%%%%%%%%%%%%%%%%%%%%%%%%%%%%%%%%%%%

\section{Results in detail and additional tables}\label{Appendix::Results}

\begin{table}[h!]  
\begin{center}
\caption{Reduced-form results in detail: Austria}\label{table::detailsAT}
\begin{tabular}{lcc}
\toprule
 & Male $(j=1)$ & Female $(j=2)$ \\ \midrule
Current shocks: &  &  \\
~~~~$\beta_{j[\omega_{jt}]}$ & $ -452.675$** & $ -302.384$** \\
 & $(   74.842)$ & $(   36.488)$ \\
~~~~$\beta_{j[\omega_{-jt}]}$ & $    9.747$ & $  -65.641$ \\
 & $(   27.454)$ & $(   51.065)$ \\
Past shocks: &  &  \\
~~~~$\beta_{j[\omega_{jt-1}]}$ & $  -71.976$** & $  -38.797$** \\
 & $(   24.014)$ & $(   12.194)$ \\
~~~~$\beta_{j[\omega_{-jt-1}]}$ & $   24.681$* & $  -20.422$ \\
 & $(   11.924)$ & $(   21.857)$ \\
Initial distribution factor: &  &  \\
~~~~$ \beta_{j[\theta_{j}]}$ & $  -61.760$* & $  -25.946$* \\
 & $(   28.589)$ & $(   13.041)$ \\
Other controls: &  &  \\
~~~~$ b_{j[\Delta y_t]}$ & $  414.117$** & $  240.520$** \\
 & $(   89.865)$ & $(   45.639)$ \\
~~~~$ b_{j[y_{t-1}]}$ & $   -2.504$ & $    5.054$* \\
 & $(    2.470)$ & $(    2.267)$ \\
~~~~$ b_{j[\Delta a_t]}$ & $   -0.094$ & $    0.304$ \\
 & $(    3.978)$ & $(    5.943)$ \\
~~~~$ b_{j[\Delta a_{t-1}]}$ & $   -1.926$ & $    2.079$ \\
 & $(    3.651)$ & $(    4.855)$ \\
~~~~$ b_{j[a_{t-1}]}$ & $    1.260$ & $  -10.466$** \\
 & $(    3.283)$ & $(    2.992)$ \\
~~~~$ b_{j[q_t]}$ & $    0.000$ & $    0.000$ \\
 & $(    0.000)$ & $(    0.000)$ \\
~~~~$ b_{j[l_{t}]}$ & $   -0.020$ & $   -0.023$ \\
 & $(    0.026)$ & $(    0.024)$ \\
~~~~$ b_{j[\Delta y_{-jt}]}$ & $ -378.371$** & $  -56.788$ \\
 & $(  121.625)$ & $(  117.612)$ \\
 &  &  \\
 Observations & $        3,170$ & $        3,170$ \\
\bottomrule
\end{tabular}
\begin{minipage}{0.7\textwidth}
\footnotesize
\textit{Notes}: The table presents the main estimates from \eqref{Eq::EstimableEquation} for Austria. Robust standard errors, clustered at the household level, in parentheses.\\
$^{**}$ significant at the 1\%; $^{*}$ significant at the 5\%.
\end{minipage}
\end{center}
\end{table}
\clearpage

\begin{table}[h!]  
\begin{center}
\caption{Reduced-form results in detail: Belgium}
\begin{tabular}{lcc}
\toprule
 & Male $(j=1)$ & Female $(j=2)$ \\ \midrule
Current shocks: &  &  \\
~~~~$\beta_{j[\omega_{jt}]}$ & $ -478.534$** & $ -241.878$** \\
 & $(   41.499)$ & $(   39.351)$ \\
~~~~$\beta_{j[\omega_{-jt}]}$ & $  -56.030$ & $  -15.671$ \\
 & $(   31.646)$ & $(   38.580)$ \\
Past shocks: &  &  \\
~~~~$\beta_{j[\omega_{jt-1}]}$ & $  -64.177$** & $   -9.725$ \\
 & $(   19.847)$ & $(   17.128)$ \\
~~~~$\beta_{j[\omega_{-jt-1}]}$ & $   -0.252$ & $   31.876$ \\
 & $(   19.947)$ & $(   20.007)$ \\
Initial distribution factor: &  &  \\
~~~~$ \beta_{j[\theta_{j}]}$ & $   -0.987$ & $   -4.098$ \\
 & $(   14.437)$ & $(   10.668)$ \\
Other controls: &  &  \\
~~~~$ b_{j[\Delta y_t]}$ & $  554.287$** & $  373.533$** \\
 & $(   60.360)$ & $(   56.852)$ \\
~~~~$ b_{j[y_{t-1}]}$ & $   -4.751$ & $   -0.936$ \\
 & $(    2.714)$ & $(    2.314)$ \\
~~~~$ b_{j[\Delta a_t]}$ & $    2.756$ & $   -9.867$* \\
 & $(    3.566)$ & $(    4.646)$ \\
~~~~$ b_{j[\Delta a_{t-1}]}$ & $   -4.302$ & $  -13.243$ \\
 & $(    3.229)$ & $(    7.697)$ \\
~~~~$ b_{j[a_{t-1}]}$ & $    2.370$ & $   -1.636$ \\
 & $(    3.439)$ & $(    2.931)$ \\
~~~~$ b_{j[q_t]}$ & $   -0.000$ & $   -0.001$ \\
 & $(    0.000)$ & $(    0.001)$ \\
~~~~$ b_{j[l_{t}]}$ & $    0.008$ & $   -0.003$ \\
 & $(    0.023)$ & $(    0.020)$ \\
~~~~$ b_{j[\Delta y_{-jt}]}$ & $ -327.231$** & $ -303.259$** \\
 & $(   86.990)$ & $(   95.953)$ \\
 &  &  \\
 Observations & $        4,115$ & $        4,115$ \\
\bottomrule
\end{tabular}
\begin{minipage}{0.7\textwidth}
\footnotesize
\textit{Notes}: The table presents the main estimates from \eqref{Eq::EstimableEquation} for Belgium. Robust standard errors, clustered at the household level, in parentheses.\\
$^{**}$ significant at the 1\%; $^{*}$ significant at the 5\%.
\end{minipage}
\end{center}
\end{table}
\clearpage

\begin{table}[h!]  
\begin{center}
\caption{Reduced-form results in detail: Bulgaria}
\begin{tabular}{lcc}
\toprule
 & Male $(j=1)$ & Female $(j=2)$ \\ \midrule
Current shocks: &  &  \\
~~~~$\beta_{j[\omega_{jt}]}$ & $  -56.414$** & $  -52.436$** \\
 & $(   15.139)$ & $(   18.045)$ \\
~~~~$\beta_{j[\omega_{-jt}]}$ & $   17.863$ & $   23.250$ \\
 & $(   14.018)$ & $(   11.892)$ \\
Past shocks: &  &  \\
~~~~$\beta_{j[\omega_{jt-1}]}$ & $    1.720$ & $   13.438$ \\
 & $(   12.451)$ & $(    9.495)$ \\
~~~~$\beta_{j[\omega_{-jt-1}]}$ & $    0.827$ & $    4.393$ \\
 & $(    5.806)$ & $(    7.503)$ \\
Initial distribution factor: &  &  \\
~~~~$ \beta_{j[\theta_{j}]}$ & $    6.855$ & $   -3.211$ \\
 & $(    4.899)$ & $(    4.136)$ \\
Other controls: &  &  \\
~~~~$ b_{j[\Delta y_t]}$ & $   58.215$** & $   44.940$** \\
 & $(   18.537)$ & $(   14.878)$ \\
~~~~$ b_{j[y_{t-1}]}$ & $   -0.315$ & $    2.149$ \\
 & $(    1.355)$ & $(    1.171)$ \\
~~~~$ b_{j[\Delta a_t]}$ & $    1.017$ & $    2.366$ \\
 & $(    3.123)$ & $(    2.146)$ \\
~~~~$ b_{j[\Delta a_{t-1}]}$ & $   -1.195$ & $    0.581$ \\
 & $(    1.123)$ & $(    1.031)$ \\
~~~~$ b_{j[a_{t-1}]}$ & $   -0.009$ & $   -3.130$ \\
 & $(    1.830)$ & $(    1.601)$ \\
~~~~$ b_{j[q_t]}$ & $    0.000$ & $    0.000$ \\
 & $(    0.000)$ & $(    0.000)$ \\
~~~~$ b_{j[l_{t}]}$ & $   -0.105$ & $   -0.068$ \\
 & $(    0.096)$ & $(    0.058)$ \\
~~~~$ b_{j[\Delta y_{-jt}]}$ & $  -79.605$* & $  -70.698$* \\
 & $(   32.036)$ & $(   29.295)$ \\
 &  &  \\
 Observations & $        3,482$ & $        3,482$ \\
\bottomrule
\end{tabular}
\begin{minipage}{0.7\textwidth}
\footnotesize
\textit{Notes}: The table presents the main estimates from \eqref{Eq::EstimableEquation} for Bulgaria. Robust standard errors, clustered at the household level, in parentheses.\\
$^{**}$ significant at the 1\%; $^{*}$ significant at the 5\%.
\end{minipage}
\end{center}
\end{table}
\clearpage

\begin{table}[h!]  
\begin{center}
\caption{Reduced-form results in detail: Cyprus}
\begin{tabular}{lcc}
\toprule
 & Male $(j=1)$ & Female $(j=2)$ \\ \midrule
Current shocks: &  &  \\
~~~~$\beta_{j[\omega_{jt}]}$ & $ -403.703$** & $ -291.287$** \\
 & $(   55.895)$ & $(   44.750)$ \\
~~~~$\beta_{j[\omega_{-jt}]}$ & $   68.699$ & $    9.439$ \\
 & $(   53.996)$ & $(   52.258)$ \\
Past shocks: &  &  \\
~~~~$\beta_{j[\omega_{jt-1}]}$ & $   -8.436$ & $    2.114$ \\
 & $(   28.992)$ & $(   24.351)$ \\
~~~~$\beta_{j[\omega_{-jt-1}]}$ & $    0.715$ & $   47.612$* \\
 & $(   18.569)$ & $(   20.141)$ \\
Initial distribution factor: &  &  \\
~~~~$ \beta_{j[\theta_{j}]}$ & $   11.905$ & $    7.293$ \\
 & $(   14.275)$ & $(    9.990)$ \\
Other controls: &  &  \\
~~~~$ b_{j[\Delta y_t]}$ & $  259.424$** & $  121.027$** \\
 & $(   49.177)$ & $(   45.325)$ \\
~~~~$ b_{j[y_{t-1}]}$ & $    1.018$ & $   -2.972$ \\
 & $(    1.327)$ & $(    2.500)$ \\
~~~~$ b_{j[\Delta a_t]}$ & $   -6.790$ & $    4.563$ \\
 & $(    4.960)$ & $(    4.805)$ \\
~~~~$ b_{j[\Delta a_{t-1}]}$ & $   -1.670$ & $    5.026$ \\
 & $(    5.200)$ & $(    5.859)$ \\
~~~~$ b_{j[a_{t-1}]}$ & $   -3.574$* & $    0.799$ \\
 & $(    1.739)$ & $(    3.089)$ \\
~~~~$ b_{j[q_t]}$ & $    0.000$ & $   -0.000$ \\
 & $(    0.000)$ & $(    0.000)$ \\
~~~~$ b_{j[l_{t}]}$ & $   -0.119$** & $   -0.013$ \\
 & $(    0.043)$ & $(    0.028)$ \\
~~~~$ b_{j[\Delta y_{-jt}]}$ & $ -135.305$ & $   20.969$ \\
 & $(  133.786)$ & $(  127.748)$ \\
 &  &  \\
 Observations & $        3,541$ & $        3,541$ \\
\bottomrule
\end{tabular}
\begin{minipage}{0.7\textwidth}
\footnotesize
\textit{Notes}: The table presents the main estimates from \eqref{Eq::EstimableEquation} for Cyprus. Robust standard errors, clustered at the household level, in parentheses.\\
$^{**}$ significant at the 1\%; $^{*}$ significant at the 5\%.
\end{minipage}
\end{center}
\end{table}
\clearpage

\begin{table}[h!]  
\begin{center}
\caption{Reduced-form results in detail: Czech Republic}
\begin{tabular}{lcc}
\toprule
 & Male $(j=1)$ & Female $(j=2)$ \\ \midrule
Current shocks: &  &  \\
~~~~$\beta_{j[\omega_{jt}]}$ & $ -378.429$** & $  -94.913$** \\
 & $(   22.330)$ & $(   21.462)$ \\
~~~~$\beta_{j[\omega_{-jt}]}$ & $  100.325$** & $    9.562$ \\
 & $(   22.118)$ & $(   37.537)$ \\
Past shocks: &  &  \\
~~~~$\beta_{j[\omega_{jt-1}]}$ & $    3.680$ & $  -86.776$** \\
 & $(   14.325)$ & $(   29.646)$ \\
~~~~$\beta_{j[\omega_{-jt-1}]}$ & $    5.749$ & $   -0.642$ \\
 & $(    8.669)$ & $(   15.298)$ \\
Initial distribution factor: &  &  \\
~~~~$ \beta_{j[\theta_{j}]}$ & $   11.672$ & $  -14.947$ \\
 & $(    8.475)$ & $(    9.186)$ \\
Other controls: &  &  \\
~~~~$ b_{j[\Delta y_t]}$ & $  281.483$** & $  125.576$** \\
 & $(   29.092)$ & $(   45.772)$ \\
~~~~$ b_{j[y_{t-1}]}$ & $   -1.434$ & $    0.024$ \\
 & $(    0.771)$ & $(    1.440)$ \\
~~~~$ b_{j[\Delta a_t]}$ & $   -0.794$ & $    1.126$ \\
 & $(    3.136)$ & $(    5.556)$ \\
~~~~$ b_{j[\Delta a_{t-1}]}$ & $    3.845$ & $   -3.180$ \\
 & $(    2.738)$ & $(    5.464)$ \\
~~~~$ b_{j[a_{t-1}]}$ & $    0.382$ & $   -0.574$ \\
 & $(    1.564)$ & $(    2.836)$ \\
~~~~$ b_{j[q_t]}$ & $    0.000$ & $   -0.001$ \\
 & $(    0.001)$ & $(    0.001)$ \\
~~~~$ b_{j[l_{t}]}$ & $   -0.118$** & $   -0.055$* \\
 & $(    0.025)$ & $(    0.027)$ \\
~~~~$ b_{j[\Delta y_{-jt}]}$ & $ -408.999$** & $ -160.652$ \\
 & $(   57.579)$ & $(   86.664)$ \\
 &  &  \\
 Observations & $        6,069$ & $        6,069$ \\ 
\bottomrule
\end{tabular}
\begin{minipage}{0.7\textwidth}
\footnotesize
\textit{Notes}: The table presents the main estimates from \eqref{Eq::EstimableEquation} for the Czech Republic. Robust standard errors, clustered at the household level, in parentheses.\\
$^{**}$ significant at the 1\%; $^{*}$ significant at the 5\%.
\end{minipage}
\end{center}
\end{table}
\clearpage

\begin{table}[h!]  
\begin{center}
\caption{Reduced-form results in detail: France}
\begin{tabular}{lcc}
\toprule
 & Male $(j=1)$ & Female $(j=2)$ \\ \midrule
Current shocks: &  &  \\
~~~~$\beta_{j[\omega_{jt}]}$ & $ -667.063$** & $ -170.289$* \\
 & $(   80.424)$ & $(   74.558)$ \\
~~~~$\beta_{j[\omega_{-jt}]}$ & $   61.794$ & $   42.654$ \\
 & $(   39.191)$ & $(   79.116)$ \\
Past shocks: &  &  \\
~~~~$\beta_{j[\omega_{jt-1}]}$ & $  -21.885$ & $  -76.676$** \\
 & $(   33.500)$ & $(   18.283)$ \\
~~~~$\beta_{j[\omega_{-jt-1}]}$ & $  -36.392$* & $  -55.618$ \\
 & $(   17.495)$ & $(   37.987)$ \\
Initial distribution factor: &  &  \\
~~~~$ \beta_{j[\theta_{j}]}$ & $   13.192$ & $   -8.423$ \\
 & $(   18.275)$ & $(   13.450)$ \\
Other controls: &  &  \\
~~~~$ b_{j[\Delta y_t]}$ & $  256.434$** & $  161.393$** \\
 & $(   57.037)$ & $(   60.432)$ \\
~~~~$ b_{j[y_{t-1}]}$ & $    1.505$ & $    4.823$ \\
 & $(    2.756)$ & $(    5.072)$ \\
~~~~$ b_{j[\Delta a_t]}$ & $    3.476$ & $    7.500$ \\
 & $(    8.405)$ & $(    8.357)$ \\
~~~~$ b_{j[\Delta a_{t-1}]}$ & $    2.629$ & $    6.586$ \\
 & $(    5.239)$ & $(    7.332)$ \\
~~~~$ b_{j[a_{t-1}]}$ & $   -3.950$ & $   -7.705$ \\
 & $(    3.599)$ & $(    6.429)$ \\
~~~~$ b_{j[q_t]}$ & $    0.000$ & $    0.000$ \\
 & $(    0.000)$ & $(    0.000)$ \\
~~~~$ b_{j[l_{t}]}$ & $   -0.079$* & $    0.024$ \\
 & $(    0.033)$ & $(    0.047)$ \\
~~~~$ b_{j[\Delta y_{-jt}]}$ & $ -417.021$* & $ -303.035$ \\
 & $(  195.314)$ & $(  169.156)$ \\
 &  &  \\
 Observations & $        2,292$ & $        2,292$ \\
\bottomrule
\end{tabular}
\begin{minipage}{0.7\textwidth}
\footnotesize
\textit{Notes}: The table presents the main estimates from \eqref{Eq::EstimableEquation} for France. Robust standard errors, clustered at the household level, in parentheses.\\
$^{**}$ significant at the 1\%; $^{*}$ significant at the 5\%.
\end{minipage}
\end{center}
\end{table}
\clearpage

\begin{table}[h!]  
\begin{center}
\caption{Reduced-form results in detail: Greece}
\begin{tabular}{lcc}
\toprule
 & Male $(j=1)$ & Female $(j=2)$ \\ \midrule
Current shocks: &  &  \\
~~~~$\beta_{j[\omega_{jt}]}$ & $ -542.965$** & $ -431.505$** \\
 & $(   44.055)$ & $(   53.997)$ \\
~~~~$\beta_{j[\omega_{-jt}]}$ & $  -42.414$ & $  -86.552$ \\
 & $(   66.442)$ & $(   50.633)$ \\
Past shocks: &  &  \\
~~~~$\beta_{j[\omega_{jt-1}]}$ & $  -58.623$** & $  -91.279$** \\
 & $(   21.994)$ & $(   33.654)$ \\
~~~~$\beta_{j[\omega_{-jt-1}]}$ & $   -9.576$ & $   25.017$ \\
 & $(   22.384)$ & $(   23.939)$ \\
Initial distribution factor: &  &  \\
~~~~$ \beta_{j[\theta_{j}]}$ & $   28.972$* & $   -5.498$ \\
 & $(   14.236)$ & $(    9.268)$ \\
Other controls: &  &  \\
~~~~$ b_{j[\Delta y_t]}$ & $  690.163$** & $  501.737$** \\
 & $(   66.403)$ & $(  115.222)$ \\
~~~~$ b_{j[y_{t-1}]}$ & $   -1.775$ & $   -4.055$** \\
 & $(    1.513)$ & $(    1.253)$ \\
~~~~$ b_{j[\Delta a_t]}$ & $   -2.117$ & $    0.392$ \\
 & $(    5.110)$ & $(    4.654)$ \\
~~~~$ b_{j[\Delta a_{t-1}]}$ & $   -7.358$** & $    1.765$ \\
 & $(    2.560)$ & $(    5.300)$ \\
~~~~$ b_{j[a_{t-1}]}$ & $    1.472$ & $    5.006$** \\
 & $(    1.819)$ & $(    1.608)$ \\
~~~~$ b_{j[q_t]}$ & $    0.000$ & $   -0.000$ \\
 & $(    0.000)$ & $(    0.001)$ \\
~~~~$ b_{j[l_{t}]}$ & $   -0.107$* & $   -0.085$* \\
 & $(    0.043)$ & $(    0.041)$ \\
~~~~$ b_{j[\Delta y_{-jt}]}$ & $ -467.728$* & $ -375.196$** \\
 & $(  224.454)$ & $(  113.986)$ \\
 &  &  \\
 Observations & $        3,095$ & $        3,095$ \\
\bottomrule
\end{tabular}
\begin{minipage}{0.7\textwidth}
\footnotesize
\textit{Notes}: The table presents the main estimates from \eqref{Eq::EstimableEquation} for Greece. Robust standard errors, clustered at the household level, in parentheses.\\
$^{**}$ significant at the 1\%; $^{*}$ significant at the 5\%.
\end{minipage}
\end{center}
\end{table}
\clearpage

\begin{table}[h!]  
\begin{center}
\caption{Reduced-form results in detail: Hungary}
\begin{tabular}{lcc}
\toprule
 & Male $(j=1)$ & Female $(j=2)$ \\ \midrule
Current shocks: &  &  \\
~~~~$\beta_{j[\omega_{jt}]}$ & $ -193.258$** & $ -297.211$** \\
 & $(   37.294)$ & $(   45.953)$ \\
~~~~$\beta_{j[\omega_{-jt}]}$ & $   83.943$* & $   17.680$ \\
 & $(   38.135)$ & $(   48.021)$ \\
Past shocks: &  &  \\
~~~~$\beta_{j[\omega_{jt-1}]}$ & $  -46.345$* & $  -44.948$ \\
 & $(   20.986)$ & $(   31.001)$ \\
~~~~$\beta_{j[\omega_{-jt-1}]}$ & $   24.302$ & $   42.069$* \\
 & $(   18.651)$ & $(   21.154)$ \\
Initial distribution factor: &  &  \\
~~~~$ \beta_{j[\theta_{j}]}$ & $   -7.415$ & $    1.293$ \\
 & $(   13.651)$ & $(    9.892)$ \\
Other controls: &  &  \\
~~~~$ b_{j[\Delta y_t]}$ & $  168.520$** & $  273.322$** \\
 & $(   40.067)$ & $(   55.218)$ \\
~~~~$ b_{j[y_{t-1}]}$ & $   -1.434$ & $    3.323$ \\
 & $(    1.616)$ & $(    2.278)$ \\
~~~~$ b_{j[\Delta a_t]}$ & $   -7.299$* & $    8.370$ \\
 & $(    2.994)$ & $(    6.914)$ \\
~~~~$ b_{j[\Delta a_{t-1}]}$ & $   -8.258$ & $   -0.005$ \\
 & $(    4.926)$ & $(    6.080)$ \\
~~~~$ b_{j[a_{t-1}]}$ & $    1.306$ & $   -5.808$ \\
 & $(    2.213)$ & $(    3.034)$ \\
~~~~$ b_{j[q_t]}$ & $    0.003$** & $    0.001$ \\
 & $(    0.001)$ & $(    0.002)$ \\
~~~~$ b_{j[l_{t}]}$ & $   -0.080$** & $   -0.023$ \\
 & $(    0.029)$ & $(    0.028)$ \\
~~~~$ b_{j[\Delta y_{-jt}]}$ & $ -304.937$** & $ -278.422$* \\
 & $(  102.496)$ & $(  112.147)$ \\
 &  &  \\
 Observations & $        2,721$ & $        2,721$ \\
\bottomrule
\end{tabular}
\begin{minipage}{0.7\textwidth}
\footnotesize
\textit{Notes}: The table presents the main estimates from \eqref{Eq::EstimableEquation} for Hungary. Robust standard errors, clustered at the household level, in parentheses.\\
$^{**}$ significant at the 1\%; $^{*}$ significant at the 5\%.
\end{minipage}
\end{center}
\end{table}
\clearpage

\begin{table}[h!]  
\begin{center}
\caption{Reduced-form results in detail: Italy}
\begin{tabular}{lcc}
\toprule
 & Male $(j=1)$ & Female $(j=2)$ \\ \midrule
Current shocks: &  &  \\
~~~~$\beta_{j[\omega_{jt}]}$ & $ -386.510$** & $ -483.033$** \\
 & $(   37.361)$ & $(   24.586)$ \\
~~~~$\beta_{j[\omega_{-jt}]}$ & $  -29.169$ & $  -35.228$ \\
 & $(   29.897)$ & $(   33.531)$ \\
Past shocks: &  &  \\
~~~~$\beta_{j[\omega_{jt-1}]}$ & $  -49.644$** & $  -59.728$** \\
 & $(   18.582)$ & $(   16.615)$ \\
~~~~$\beta_{j[\omega_{-jt-1}]}$ & $  -19.837$ & $  -18.546$ \\
 & $(   12.417)$ & $(   16.554)$ \\
Initial distribution factor: &  &  \\
~~~~$ \beta_{j[\theta_{j}]}$ & $  -20.588$ & $    0.306$ \\
 & $(   13.931)$ & $(    8.600)$ \\
Other controls: &  &  \\
~~~~$ b_{j[\Delta y_t]}$ & $  226.281$** & $  237.656$** \\
 & $(   31.581)$ & $(   29.037)$ \\
~~~~$ b_{j[y_{t-1}]}$ & $   -2.045$ & $   -1.298$ \\
 & $(    1.943)$ & $(    1.652)$ \\
~~~~$ b_{j[\Delta a_t]}$ & $    6.192$ & $   -0.265$ \\
 & $(    4.494)$ & $(    3.767)$ \\
~~~~$ b_{j[\Delta a_{t-1}]}$ & $    3.716$ & $   -2.600$ \\
 & $(    3.052)$ & $(    2.416)$ \\
~~~~$ b_{j[a_{t-1}]}$ & $    1.396$ & $   -1.467$ \\
 & $(    2.494)$ & $(    2.088)$ \\
~~~~$ b_{j[q_t]}$ & $    0.000$ & $   -0.000$ \\
 & $(    0.000)$ & $(    0.000)$ \\
~~~~$ b_{j[l_{t}]}$ & $   -0.013$ & $    0.004$ \\
 & $(    0.027)$ & $(    0.017)$ \\
~~~~$ b_{j[\Delta y_{-jt}]}$ & $  -74.274$ & $ -120.098$ \\
 & $(   90.997)$ & $(   72.618)$ \\
 &  &  \\
 Observations & $        6,779$ & $        6,779$ \\
\bottomrule
\end{tabular}
\begin{minipage}{0.7\textwidth}
\footnotesize
\textit{Notes}: The table presents the main estimates from \eqref{Eq::EstimableEquation} for Italy. Robust standard errors, clustered at the household level, in parentheses.\\
$^{**}$ significant at the 1\%; $^{*}$ significant at the 5\%.
\end{minipage}
\end{center}
\end{table}
\clearpage

\begin{table}[h!]  
\begin{center}
\caption{Reduced-form results in detail: Luxembourg}
\begin{tabular}{lcc}
\toprule
 & Male $(j=1)$ & Female $(j=2)$ \\ \midrule
Current shocks: &  &  \\
~~~~$\beta_{j[\omega_{jt}]}$ & $ -399.093$** & $ -242.285$** \\
 & $(   42.363)$ & $(   29.881)$ \\
~~~~$\beta_{j[\omega_{-jt}]}$ & $  -24.459$ & $  -55.821$ \\
 & $(   32.980)$ & $(   49.678)$ \\
Past shocks: &  &  \\
~~~~$\beta_{j[\omega_{jt-1}]}$ & $  -65.318$* & $  -56.195$** \\
 & $(   27.445)$ & $(   17.109)$ \\
~~~~$\beta_{j[\omega_{-jt-1}]}$ & $    6.640$ & $   42.829$ \\
 & $(   11.104)$ & $(   26.896)$ \\
Initial distribution factor: &  &  \\
~~~~$ \beta_{j[\theta_{j}]}$ & $    9.356$ & $  -17.283$ \\
 & $(   13.065)$ & $(   16.049)$ \\
Other controls: &  &  \\
~~~~$ b_{j[\Delta y_t]}$ & $  371.655$** & $  345.092$** \\
 & $(   59.634)$ & $(   57.053)$ \\
~~~~$ b_{j[y_{t-1}]}$ & $   -0.203$ & $   -0.878$ \\
 & $(    1.350)$ & $(    2.020)$ \\
~~~~$ b_{j[\Delta a_t]}$ & $   -1.779$ & $   -3.672$ \\
 & $(    2.085)$ & $(    2.998)$ \\
~~~~$ b_{j[\Delta a_{t-1}]}$ & $    1.425$ & $   -0.050$ \\
 & $(    1.881)$ & $(    3.362)$ \\
~~~~$ b_{j[a_{t-1}]}$ & $   -1.875$ & $   -0.351$ \\
 & $(    1.875)$ & $(    2.592)$ \\
~~~~$ b_{j[q_t]}$ & $    0.000$ & $   -0.000$ \\
 & $(    0.000)$ & $(    0.000)$ \\
~~~~$ b_{j[l_{t}]}$ & $   -0.005$ & $    0.005$ \\
 & $(    0.026)$ & $(    0.032)$ \\
~~~~$ b_{j[\Delta y_{-jt}]}$ & $  -52.277$ & $ -163.114$ \\
 & $(  103.152)$ & $(  120.358)$ \\
 &  &  \\
 Observations & $        3,768$ & $        3,768$ \\
\bottomrule
\end{tabular}
\begin{minipage}{0.7\textwidth}
\footnotesize
\textit{Notes}: The table presents the main estimates from \eqref{Eq::EstimableEquation} for Luxembourg. Robust standard errors, clustered at the household level, in parentheses.\\
$^{**}$ significant at the 1\%; $^{*}$ significant at the 5\%.
\end{minipage}
\end{center}
\end{table}
\clearpage

\begin{table}[h!]  
\begin{center}
\caption{Reduced-form results in detail: Poland}
\begin{tabular}{lcc}
\toprule
 & Male $(j=1)$ & Female $(j=2)$ \\ \midrule
Current shocks: &  &  \\
~~~~$\beta_{j[\omega_{jt}]}$ & $ -302.976$** & $ -293.569$** \\
 & $(   25.203)$ & $(   53.944)$ \\
~~~~$\beta_{j[\omega_{-jt}]}$ & $   17.773$ & $   65.818$ \\
 & $(   21.028)$ & $(   55.155)$ \\
Past shocks: &  &  \\
~~~~$\beta_{j[\omega_{jt-1}]}$ & $    9.261$ & $  -75.948$** \\
 & $(   18.545)$ & $(   22.232)$ \\
~~~~$\beta_{j[\omega_{-jt-1}]}$ & $    1.093$ & $  -19.034$ \\
 & $(    9.865)$ & $(   20.032)$ \\
Initial distribution factor: &  &  \\
~~~~$ \beta_{j[\theta_{j}]}$ & $  -10.659$ & $   -2.326$ \\
 & $(    7.451)$ & $(   12.028)$ \\
Other controls: &  &  \\
~~~~$ b_{j[\Delta y_t]}$ & $  300.081$** & $  286.977$** \\
 & $(   36.133)$ & $(   63.588)$ \\
~~~~$ b_{j[y_{t-1}]}$ & $    0.714$ & $    1.854$ \\
 & $(    1.253)$ & $(    1.532)$ \\
~~~~$ b_{j[\Delta a_t]}$ & $   -2.718$ & $   -6.104$* \\
 & $(    2.466)$ & $(    2.374)$ \\
~~~~$ b_{j[\Delta a_{t-1}]}$ & $   -2.515$ & $   -1.822$ \\
 & $(    1.832)$ & $(    2.485)$ \\
~~~~$ b_{j[a_{t-1}]}$ & $   -3.130$ & $   -5.441$** \\
 & $(    1.709)$ & $(    1.870)$ \\
~~~~$ b_{j[q_t]}$ & $    0.000$ & $   -0.002$* \\
 & $(    0.001)$ & $(    0.001)$ \\
~~~~$ b_{j[l_{t}]}$ & $   -0.063$** & $    0.000$ \\
 & $(    0.018)$ & $(    0.038)$ \\
~~~~$ b_{j[\Delta y_{-jt}]}$ & $ -285.291$** & $ -356.535$** \\
 & $(   72.620)$ & $(  116.792)$ \\
 &  &  \\
 Observations & $        8,351$ & $        8,351$ \\
\bottomrule
\end{tabular}
\begin{minipage}{0.7\textwidth}
\footnotesize
\textit{Notes}: The table presents the main estimates from \eqref{Eq::EstimableEquation} for Poland. Robust standard errors, clustered at the household level, in parentheses.\\
$^{**}$ significant at the 1\%; $^{*}$ significant at the 5\%.
\end{minipage}
\end{center}
\end{table}
\clearpage

\begin{table}[h!]  
\begin{center}
\caption{Reduced-form results in detail: Romania}
\begin{tabular}{lcc}
\toprule
 & Male $(j=1)$ & Female $(j=2)$ \\ \midrule
Current shocks: &  &  \\
~~~~$\beta_{j[\omega_{jt}]}$ & $ -221.465$** & $  -83.205$** \\
 & $(   33.736)$ & $(   29.129)$ \\
~~~~$\beta_{j[\omega_{-jt}]}$ & $   88.348$** & $   93.875$** \\
 & $(   31.007)$ & $(   28.064)$ \\
Past shocks: &  &  \\
~~~~$\beta_{j[\omega_{jt-1}]}$ & $  -52.199$* & $  -87.949$** \\
 & $(   25.226)$ & $(   27.228)$ \\
~~~~$\beta_{j[\omega_{-jt-1}]}$ & $   33.043$ & $   24.572$ \\
 & $(   20.718)$ & $(   16.012)$ \\
Initial distribution factor: &  &  \\
~~~~$ \beta_{j[\theta_{j}]}$ & $   12.958$ & $    2.314$ \\
 & $(    7.870)$ & $(    3.440)$ \\
Other controls: &  &  \\
~~~~$ b_{j[\Delta y_t]}$ & $  154.280$** & $   61.551$* \\
 & $(   33.185)$ & $(   30.306)$ \\
~~~~$ b_{j[y_{t-1}]}$ & $   -1.884$* & $    0.027$ \\
 & $(    0.770)$ & $(    0.842)$ \\
~~~~$ b_{j[\Delta a_t]}$ & $    0.832$ & $   -0.251$ \\
 & $(    1.581)$ & $(    0.962)$ \\
~~~~$ b_{j[\Delta a_{t-1}]}$ & $   -0.743$ & $    0.458$ \\
 & $(    1.502)$ & $(    1.086)$ \\
~~~~$ b_{j[a_{t-1}]}$ & $    2.458$ & $   -0.297$ \\
 & $(    1.481)$ & $(    1.824)$ \\
~~~~$ b_{j[q_t]}$ & $    0.001$ & $    0.002$ \\
 & $(    0.001)$ & $(    0.000)$ \\
~~~~$ b_{j[l_{t}]}$ & $   -0.145$** & $   -0.098$** \\
 & $(    0.047)$ & $(    0.027)$ \\
~~~~$ b_{j[\Delta y_{-jt}]}$ & $ -217.952$** & $ -168.858$* \\
 & $(   84.265)$ & $(   69.284)$ \\
 &  &  \\
 Observations & $        4,912$ & $        4,912$ \\
\bottomrule
\end{tabular}
\begin{minipage}{0.7\textwidth}
\footnotesize
\textit{Notes}: The table presents the main estimates from \eqref{Eq::EstimableEquation} for Romania. Robust standard errors, clustered at the household level, in parentheses.\\
$^{**}$ significant at the 1\%; $^{*}$ significant at the 5\%.
\end{minipage}
\end{center}
\end{table}
\clearpage

\begin{table}[h!]  
\begin{center}
\caption{Reduced-form results in detail: Slovakia}
\begin{tabular}{lcc}
\toprule
 & Male $(j=1)$ & Female $(j=2)$ \\ \midrule
Current shocks: &  &  \\
~~~~$\beta_{j[\omega_{jt}]}$ & $ -213.993$** & $ -210.597$** \\
 & $(   24.213)$ & $(   40.103)$ \\
~~~~$\beta_{j[\omega_{-jt}]}$ & $  -61.802$ & $   57.362$ \\
 & $(   66.259)$ & $(   40.291)$ \\
Past shocks: &  &  \\
~~~~$\beta_{j[\omega_{jt-1}]}$ & $  -11.121$ & $   -0.952$ \\
 & $(   11.180)$ & $(   16.726)$ \\
~~~~$\beta_{j[\omega_{-jt-1}]}$ & $    3.090$ & $   24.163$ \\
 & $(   14.787)$ & $(   14.178)$ \\
Initial distribution factor: &  &  \\
~~~~$ \beta_{j[\theta_{j}]}$ & $   10.155$ & $    4.841$ \\
 & $(    9.623)$ & $(    6.302)$ \\
Other controls: &  &  \\
~~~~$ b_{j[\Delta y_t]}$ & $  204.007$** & $  200.054$** \\
 & $(   26.815)$ & $(   33.676)$ \\
~~~~$ b_{j[y_{t-1}]}$ & $   -0.922$ & $   -1.610$ \\
 & $(    0.951)$ & $(    1.200)$ \\
~~~~$ b_{j[\Delta a_t]}$ & $    2.293$ & $    0.396$ \\
 & $(    3.049)$ & $(    2.885)$ \\
~~~~$ b_{j[\Delta a_{t-1}]}$ & $   -3.921$* & $   -0.597$ \\
 & $(    1.626)$ & $(    1.778)$ \\
~~~~$ b_{j[a_{t-1}]}$ & $   -0.726$ & $    0.330$ \\
 & $(    1.357)$ & $(    1.650)$ \\
~~~~$ b_{j[q_t]}$ & $    0.000$ & $   -0.000$ \\
 & $(    0.000)$ & $(    0.000)$ \\
~~~~$ b_{j[l_{t}]}$ & $   -0.024$ & $   -0.065$** \\
 & $(    0.069)$ & $(    0.021)$ \\
~~~~$ b_{j[\Delta y_{-jt}]}$ & $  -99.139$ & $ -280.193$** \\
 & $(   88.000)$ & $(   74.364)$ \\
 &  &  \\
 Observations & $        4,440$ & $        4,440$ \\
\bottomrule
\end{tabular}
\begin{minipage}{0.7\textwidth}
\footnotesize
\textit{Notes}: The table presents the main estimates from \eqref{Eq::EstimableEquation} for Slovakia. Robust standard errors, clustered at the household level, in parentheses.\\
$^{**}$ significant at the 1\%; $^{*}$ significant at the 5\%.
\end{minipage}
\end{center}
\end{table}
\clearpage

\begin{table}[h!]  
\begin{center}
\caption{Reduced-form results in detail: Spain}
\begin{tabular}{lcc}
\toprule
 & Male $(j=1)$ & Female $(j=2)$ \\ \midrule
Current shocks: &  &  \\
~~~~$\beta_{j[\omega_{jt}]}$ & $ -339.853$** & $ -286.678$** \\
 & $(   60.816)$ & $(   44.055)$ \\
~~~~$\beta_{j[\omega_{-jt}]}$ & $  -29.861$ & $   41.527$ \\
 & $(   41.378)$ & $(   49.854)$ \\
Past shocks: &  &  \\
~~~~$\beta_{j[\omega_{jt-1}]}$ & $   14.853$ & $  -57.977$* \\
 & $(   30.859)$ & $(   29.336)$ \\
~~~~$\beta_{j[\omega_{-jt-1}]}$ & $  -72.881$ & $   25.159$ \\
 & $(   39.266)$ & $(   17.603)$ \\
Initial distribution factor: &  &  \\
~~~~$ \beta_{j[\theta_{j}]}$ & $  -33.031$* & $   -7.781$ \\
 & $(   16.736)$ & $(   24.293)$ \\
Other controls: &  &  \\
~~~~$ b_{j[\Delta y_t]}$ & $  382.664$** & $  161.978$** \\
 & $(   49.290)$ & $(   55.959)$ \\
~~~~$ b_{j[y_{t-1}]}$ & $   -0.842$ & $   -6.124$ \\
 & $(    2.693)$ & $(    4.327)$ \\
~~~~$ b_{j[\Delta a_t]}$ & $   -0.791$ & $    2.197$ \\
 & $(    5.242)$ & $(    7.990)$ \\
~~~~$ b_{j[\Delta a_{t-1}]}$ & $   -4.544$ & $  -17.524$ \\
 & $(    5.882)$ & $(   11.524)$ \\
~~~~$ b_{j[a_{t-1}]}$ & $   -1.677$ & $    3.833$ \\
 & $(    3.470)$ & $(    5.605)$ \\
~~~~$ b_{j[q_t]}$ & $   -0.001$ & $   -0.000$ \\
 & $(    0.000)$ & $(    0.001)$ \\
~~~~$ b_{j[l_{t}]}$ & $   -0.083$** & $   -0.037$ \\
 & $(    0.030)$ & $(    0.029)$ \\
~~~~$ b_{j[\Delta y_{-jt}]}$ & $ -295.603$* & $ -263.531$* \\
 & $(  133.260)$ & $(  122.562)$ \\
 &  &  \\
 Observations & $        5,136$ & $        5,136$ \\
\bottomrule
\end{tabular}
\begin{minipage}{0.7\textwidth}
\footnotesize
\textit{Notes}: The table presents the main estimates from \eqref{Eq::EstimableEquation} for Spain. Robust standard errors, clustered at the household level, in parentheses.\\
$^{**}$ significant at the 1\%; $^{*}$ significant at the 5\%.
\end{minipage}
\end{center}
\end{table}
\clearpage

\begin{table}[h!]  
\begin{center}
\caption{Reduced-form results in detail: United Kingdom}\label{table::detailsUK}
\begin{tabular}{lcc}
\toprule
 & Male $(j=1)$ & Female $(j=2)$ \\ \midrule
Current shocks: &  &  \\
~~~~$\beta_{j[\omega_{jt}]}$ & $ -535.545$** & $ -163.441$** \\
 & $(   46.744)$ & $(   34.763)$ \\
~~~~$\beta_{j[\omega_{-jt}]}$ & $   32.395$ & $    1.784$ \\
 & $(   33.649)$ & $(   44.479)$ \\
Past shocks: &  &  \\
~~~~$\beta_{j[\omega_{jt-1}]}$ & $  -46.425$ & $   -8.152$ \\
 & $(   37.751)$ & $(   26.126)$ \\
~~~~$\beta_{j[\omega_{-jt-1}]}$ & $   -0.302$ & $   15.219$ \\
 & $(   19.348)$ & $(   17.226)$ \\
Initial distribution factor: &  &  \\
~~~~$ \beta_{j[\theta_{j}]}$ & $   24.471$ & $   -9.798$ \\
 & $(   13.659)$ & $(   14.756)$ \\
Other controls: &  &  \\
~~~~$ b_{j[\Delta y_t]}$ & $  445.839$** & $  248.525$** \\
 & $(   69.291)$ & $(   51.345)$ \\
~~~~$ b_{j[y_{t-1}]}$ & $   -1.253$ & $   -1.599$ \\
 & $(    1.238)$ & $(    1.522)$ \\
~~~~$ b_{j[\Delta a_t]}$ & $   -0.724$ & $   -0.442$ \\
 & $(    1.746)$ & $(    2.521)$ \\
~~~~$ b_{j[\Delta a_{t-1}]}$ & $    0.983$ & $   -2.278$ \\
 & $(    1.799)$ & $(    2.151)$ \\
~~~~$ b_{j[a_{t-1}]}$ & $   -0.490$ & $    0.048$ \\
 & $(    2.038)$ & $(    2.344)$ \\
~~~~$ b_{j[q_t]}$ & $   -0.000$ & $    0.000$ \\
 & $(    0.000)$ & $(    0.000)$ \\
~~~~$ b_{j[l_{t}]}$ & $   -0.019$ & $   -0.039$ \\
 & $(    0.027)$ & $(    0.028)$ \\
~~~~$ b_{j[\Delta y_{-jt}]}$ & $ -247.263$* & $ -221.351$** \\
 & $(  121.251)$ & $(   82.500)$ \\
 &  &  \\
 Observations & $        3,845$ & $        3,845$ \\
\bottomrule
\end{tabular}
\begin{minipage}{0.7\textwidth}
\footnotesize
\textit{Notes}: The table presents the main estimates from \eqref{Eq::EstimableEquation} for the UK. Robust standard errors, clustered at the household level, in parentheses.\\
$^{**}$ significant at the 1\%; $^{*}$ significant at the 5\%.
\end{minipage}
\end{center}
\end{table}
\clearpage

\begin{sidewaystable}[h!]  
\begin{center}
\caption{Estimates of the Pareto weight elasticities}\label{table::elasticities}
\begin{tabular}{lcccccccccc}
\toprule
                                                      & \mcl{2}{c}{Austria}           &\mcl{2}{c}{Belgium}         & \mcl{2}{c}{Bulgaria}        & \mcl{2}{c}{Cyprus}          & \mcl{2}{c}{Czech Republic}      \\ \cmidrule{2-11}
                                                      & Male            & Female          & Male            & Female          & Male            & Female          & Male            & Female          & Male            & Female          \\
						      & $(j=1)$ & $(j=2)$ & $(j=1)$ & $(j=2)$ & $(j=1)$ & $(j=2)$ & $(j=1)$ & $(j=2)$ & $(j=1)$ & $(j=2)$ \\ \midrule
\mcl{11}{l}{Elasticity of Pareto weight w.r.t. current shocks:}  \\
$ e_{[\mu_{j}, w_{j}]}$                                 & $    1.145$**   & $    1.452$**   & $    1.204$**   & $    1.122$**   & $    1.068$**   & $    1.091$**   & $    1.620$     & $    2.326$     & $    1.061$**   & $    1.069$**   \\
                                                      & $(    0.051)$   & $(    0.502)$   & $(    0.070)$   & $(    0.047)$   & $(    0.060)$   & $(    0.059)$   & $(    0.909)$   & $(    4.302)$   & $(    0.020)$   & $(    0.063)$   \\
$ e_{[\mu_{j}, w_{-j}]}$                                & $    0.003$     & $    0.078$     & $    0.033$     & $    0.017$     & $   -0.023$     & $   -0.052$**   & $   -0.057$     & $   -0.010$     & $   -0.047$**   & $   -0.022$     \\
                                                      & $(    0.011)$   & $(    0.146)$   & $(    0.023)$   & $(    0.027)$   & $(    0.027)$   & $(    0.018)$   & $(    0.047)$   & $(    0.229)$   & $(    0.006)$   & $(    0.031)$   \\
\mcl{11}{l}{Elasticity of Pareto weight w.r.t. past shocks:}     \\
$ e_{[\mu_{j}, \mu_{jL}]} \times   e_{[\mu_{j}, w_{j}]}$  & $    0.026$*    & $    0.054$     & $    0.033$*    & $    0.013$     & $    0.016$     & $   -0.039$     & $    0.033$     & $   -0.088$     & $   -0.003$     & $    0.086$     \\
                                                      & $(    0.011)$   & $(    0.058)$   & $(    0.014)$   & $(    0.011)$   & $(    0.031)$   & $(    0.027)$   & $(    0.082)$   & $(    0.293)$   & $(    0.006)$   & $(    0.055)$   \\
$ e_{[\mu_{j}, \mu_{jL}]} \times   e_{[\mu_{j}, w_{-j}]}$ & $   -0.008$     & $    0.008$     & $    0.002$     & $   -0.015$     & $   -0.009$     & $   -0.013$     & $    0.008$     & $   -0.211$     & $    0.000$     & $   -0.002$     \\
                                                      & $(    0.005)$   & $(    0.031)$   & $(    0.010)$   & $(    0.011)$   & $(    0.013)$   & $(    0.019)$   & $(    0.036)$   & $(    0.666)$   & $(    0.003)$   & $(    0.015)$   \\
\mcl{11}{l}{Elasticity of Pareto weight w.r.t. initial distribution factor:}     \\
$ e_{[\mu_{j}, \mu_{jL}]}^t \times   e_{[\mu_{j}, \theta_{j}]}$       & $    0.025$     & $    0.421$     & $   -0.002$     & $    0.256$**   & $   -0.022$     & $    0.147$**   & $   -0.037$     & $    0.658$     & $   -0.004$     & $    0.141$**   \\
                                                                  & $(    0.014)$   & $(    0.401)$   & $(    0.007)$   & $(    0.063)$   & $(    0.017)$   & $(    0.040)$   & $(    0.056)$   & $(    1.959)$   & $(    0.003)$   & $(    0.053)$   \\
                                                      &                 &                 &                 &                 &                 &                 &                 &                 &                 &                 \\
\# obs.                                          & $        3,170$ & $        3,170$ & $        4,115$ & $        4,115$ & $        3,482$ & $        3,482$ & $        3,541$ & $        3,541$ & $        6,069$ & $        6,069$ \\
\bottomrule
\end{tabular}
\begin{minipage}{0.95\textwidth}
\footnotesize
\textit{Notes}: The table presents the structural estimates of the Pareto weight, estimated via nonlinear GMM through the system of equations \eqref{Eq::betas}. Robust standard errors, clustered at the household level, in parentheses. $^{**}$ significant at the 1\%; $^{*}$ significant at the 5\%.
\end{minipage}
\end{center}
\end{sidewaystable}
\clearpage

\begin{sidewaystable}[h!]  
\begin{center}
\caption*{Table \ref{table::elasticities} (\textit{cont.}): Estimates of the Pareto weight elasticities}
\begin{tabular}{lcccccccccc}
\toprule
                                                      & \mcl{2}{c}{France}          & \mcl{2}{c}{Greece}          & \mcl{2}{c}{Hungary}         & \mcl{2}{c}{Italy}           & \mcl{2}{c}{Luxembourg}      \\ \cmidrule{2-11}
                                                      & Male            & Female          & Male            & Female          & Male            & Female          & Male            & Female          & Male            & Female          \\
						      & $(j=1)$ & $(j=2)$ & $(j=1)$ & $(j=2)$ & $(j=1)$ & $(j=2)$ & $(j=1)$ & $(j=2)$ & $(j=1)$ & $(j=2)$ \\ \midrule
\mcl{11}{l}{Elasticity of Pareto weight w.r.t. current shocks:}  \\
$ e_{[\mu_{j}, w_{j}]}$                                 & $    1.188$**   & $    1.105$**   & $    1.064$**   & $    1.149$**   & $    1.071$**   & $    1.131$**   & $    1.876$     & $    1.768$**   & $    2.500$     & $    1.070$**   \\
                                                      & $(    0.090)$   & $(    0.101)$   & $(    0.028)$   & $(    0.049)$   & $(    0.035)$   & $(    0.055)$   & $(    1.162)$   & $(    0.437)$   & $(    2.543)$   & $(    0.037)$   \\
$ e_{[\mu_{j}, w_{-j}]}$                                & $   -0.023$**   & $    0.003$     & $   -0.008$     & $    0.037$     & $   -0.027$     & $    0.014$     & $    0.089$     & $    0.042$     & $    0.196$     & $   -0.040$*    \\
                                                      & $(    0.008)$   & $(    0.062)$   & $(    0.009)$   & $(    0.028)$   & $(    0.015)$   & $(    0.028)$   & $(    0.177)$   & $(    0.058)$   & $(    0.390)$   & $(    0.018)$   \\
\mcl{11}{l}{Elasticity of Pareto weight w.r.t. past shocks:}     \\
$ e_{[\mu_{j}, \mu_{jL}]} \times   e_{[\mu_{j}, w_{j}]}$  & $    0.007$     & $    0.061$     & $    0.015$*    & $    0.025$*    & $    0.041$**   & $    0.020$     & $    0.172$     & $    0.104$     & $    0.296$     & $    0.009$     \\
                                                      & $(    0.013)$   & $(    0.041)$   & $(    0.007)$   & $(    0.010)$   & $(    0.016)$   & $(    0.022)$   & $(    0.222)$   & $(    0.061)$   & $(    0.495)$   & $(    0.007)$   \\
$ e_{[\mu_{j}, \mu_{jL}]} \times   e_{[\mu_{j}, w_{-j}]}$ & $    0.011$     & $    0.052$     & $   -0.001$     & $   -0.002$     & $   -0.010$     & $   -0.023$     & $    0.051$     & $    0.024$     & $    0.006$     & $   -0.022$     \\
                                                      & $(    0.007)$   & $(    0.050)$   & $(    0.004)$   & $(    0.007)$   & $(    0.014)$   & $(    0.013)$   & $(    0.071)$   & $(    0.026)$   & $(    0.042)$   & $(    0.015)$   \\
\mcl{11}{l}{Elasticity of Pareto weight w.r.t. initial distribution factor:}     \\
$ e_{[\mu_{j}, \mu_{jL}]}^t \times   e_{[\mu_{j}, \theta_{j}]}$       & $   -0.006$     & $    0.123$     & $   -0.002$     & $    0.248$**   & $   -0.001$     & $    0.214$**   & $    0.021$     & $    0.354$*    & $   -0.030$     & $    0.187$**   \\
                                                                  & $(    0.007)$   & $(    0.081)$   & $(    0.003)$   & $(    0.043)$   & $(    0.009)$   & $(    0.052)$   & $(    0.034)$   & $(    0.176)$   & $(    0.071)$   & $(    0.053)$   \\
                                                      &                 &                 &                 &                 &                 &                 &                 &                 &                 &                 \\
\# obs.                                          & $        2,292$ & $        2,292$ & $        3,095$ & $        3,095$ & $        2,721$ & $        2,721$ & $        6,779$ & $        6,779$ & $        3,768$ & $        3,768$ \\
\bottomrule
\end{tabular}
\begin{minipage}{0.95\textwidth}
\footnotesize
\textit{Notes}: The table presents the structural estimates of the Pareto weight, estimated via nonlinear GMM through the system of equations \eqref{Eq::betas}. Robust standard errors, clustered at the household level, in parentheses. $^{**}$ significant at the 1\%; $^{*}$ significant at the 5\%.
\end{minipage}
\end{center}
\end{sidewaystable}
\clearpage

\begin{sidewaystable}[h!]  
\begin{center}
\caption*{Table \ref{table::elasticities} (\textit{cont.}): Estimates of the Pareto weight elasticities}
\begin{tabular}{lcccccccccc}
\toprule
                                                      & \mcl{2}{c}{Poland}          & \mcl{2}{c}{Romania}         & \mcl{2}{c}{Slovakia}        & \mcl{2}{c}{Spain}           & \mcl{2}{c}{United Kingdom}              \\ \cmidrule{2-11}
                                                      & Male            & Female          & Male            & Female          & Male            & Female          & Male            & Female          & Male            & Female          \\
						      & $(j=1)$ & $(j=2)$ & $(j=1)$ & $(j=2)$ & $(j=1)$ & $(j=2)$ & $(j=1)$ & $(j=2)$ & $(j=1)$ & $(j=2)$ \\ \midrule
\mcl{11}{l}{Elasticity of Pareto weight w.r.t. current shocks:}  \\
$ e_{[\mu_{j}, w_{j}]}$                                 & $    1.140$**   & $    1.061$**   & $    1.084$**   & $    1.007$**   & $    1.243$**   & $    1.108$**   & $    1.175$**   & $    1.166$**   & $    1.214$**   & $    1.109$**   \\
                                                      & $(    0.051)$   & $(    0.031)$   & $(    0.049)$   & $(    0.026)$   & $(    0.151)$   & $(    0.042)$   & $(    0.111)$   & $(    0.105)$   & $(    0.115)$   & $(    0.044)$   \\
$ e_{[\mu_{j}, w_{-j}]}$                                & $    0.005$     & $   -0.031$*    & $   -0.070$**   & $   -0.102$**   & $    0.042$     & $   -0.034$*    & $    0.019$     & $   -0.011$     & $   -0.022$     & $   -0.009$     \\
                                                      & $(    0.015)$   & $(    0.014)$   & $(    0.013)$   & $(    0.021)$   & $(    0.076)$   & $(    0.017)$   & $(    0.035)$   & $(    0.030)$   & $(    0.016)$   & $(    0.033)$   \\
\mcl{11}{l}{Elasticity of Pareto weight w.r.t. past shocks:}     \\
$ e_{[\mu_{j}, \mu_{jL}]} \times   e_{[\mu_{j}, w_{j}]}$  & $   -0.001$     & $    0.025$**   & $    0.023$     & $    0.089$**   & $    0.022$     & $    0.021$     & $    0.008$     & $    0.036$     & $    0.041$*    & $    0.007$     \\
                                                      & $(    0.010)$   & $(    0.008)$   & $(    0.018)$   & $(    0.029)$   & $(    0.020)$   & $(    0.017)$   & $(    0.021)$   & $(    0.027)$   & $(    0.020)$   & $(    0.019)$   \\
$ e_{[\mu_{j}, \mu_{jL}]} \times   e_{[\mu_{j}, w_{-j}]}$ & $    0.006$     & $    0.003$     & $   -0.013$     & $   -0.029$*    & $   -0.017$     & $   -0.016$     & $    0.035$     & $   -0.014$     & $    0.006$     & $   -0.020$     \\
                                                      & $(    0.007)$   & $(    0.006)$   & $(    0.014)$   & $(    0.014)$   & $(    0.015)$   & $(    0.009)$   & $(    0.032)$   & $(    0.014)$   & $(    0.011)$   & $(    0.014)$   \\
\mcl{11}{l}{Elasticity of Pareto weight w.r.t. initial distribution factor:}     \\
$ e_{[\mu_{j}, \mu_{jL}]}^t \times   e_{[\mu_{j}, \theta_{j}]}$       & $    0.014$     & $    0.144$**   & $   -0.004$     & $    0.056$**   & $   -0.007$     & $    0.156$**   & $    0.017$     & $    0.159$*    & $   -0.012$     & $    0.222$**   \\
                                                                  & $(    0.008)$   & $(    0.028)$   & $(    0.004)$   & $(    0.020)$   & $(    0.013)$   & $(    0.031)$   & $(    0.011)$   & $(    0.076)$   & $(    0.009)$   & $(    0.059)$   \\
                                                      &                 &                 &                 &                 &                 &                 &                 &                 &                 &                 \\
\# obs.                                          & $        8,351$ & $        8,351$ & $        4,912$ & $        4,912$ & $        4,440$ & $        4,440$ & $        5,136$ & $        5,136$ & $        3,845$ & $        3,845$ \\
\bottomrule
\end{tabular}
\begin{minipage}{0.96\textwidth}
\footnotesize
\textit{Notes}: The table presents the structural estimates of the Pareto weight, estimated via nonlinear GMM through the system of equations \eqref{Eq::betas}. Robust standard errors, clustered at the household level, in parentheses. $^{**}$ significant at the 1\%; $^{*}$ significant at the 5\%.
\end{minipage}
\end{center}
\end{sidewaystable}

\begin{table}[t]  
\begin{center}
\caption{Commitment test $p$-values, married couples}\label{table::test_married}
\begin{tabular}{L{2.5cm}C{2cm}C{2cm}C{2cm}C{2cm}}
\toprule
           &                          &                          & \multicolumn{2}{c}{new hypotheses}\\
\cmidrule{4-5} 
           		& ${\cal H}_{0}^\text{FC}$ & ${\cal H}_{0}^\text{NC}$ & ${\cal H}_{0}^\text{NC-new}$             & ${\cal H}_{0}^\text{LC}$   \\ 
\midrule
Austria    	&  	$<0.001$  	&  	$<0.001$  	&  	$<0.001$  	&  	$0.542$  	\\ 
Belgium    	&  	$0.025$  	&  	$0.041$  	&  	$0.015$  	&  	$0.654$  	\\
Bulgaria   	&  	$0.098$  	&  	$0.106$  	&  	$0.166$  	&  	$0.258$  	\\
Cyprus     	&  	$0.161$  	&  	$0.228$  	&  	$0.322$  	&  	$0.789$  	\\
Czech Rep. 	&  	$<0.001$  	&  	$0.006$  	&  	$0.012$  	&  	$0.682$  	\\
France     	&  	$0.002$  	&  	$0.003$  	&  	$0.005$  	&  	$0.107$  	\\
Greece     	&  	$0.002$  	&  	$<0.001$  	&  	$<0.001$  	&  	$0.081$  	\\
Hungary    	&  	$0.105$  	&  	$0.071$  	&  	$0.118$  	&  	$0.999$  	\\
Italy      		&  	$<0.001$  	&  	$<0.001$  	&  	$<0.001$  	&  	$0.343$  	\\
Luxembourg &  	$<0.001$  	&  	$<0.001$  	&  	$<0.001$  	&  	$0.481$  	\\
Poland     	&  	$0.023$  	&  	$0.012$  	&  	$0.023$  	&  	$0.717$  	\\
Romania    	&  	$<0.001$  	&  	$<0.001$  	&  	$<0.001$  	&  	$0.376$  	\\
Slovakia   	&  	$0.074$  	&  	$0.222$  	&  	$0.122$  	&  	$0.476$  	\\
Spain      	&  	$0.210$  	&  	$0.109$  	&  	$0.164$  	&  	$0.539$  	\\
UK         		&  	$0.573$  	&  	$0.579$  	&  	$0.687$  	&  	$0.587$  	\\
\bottomrule
\end{tabular}
\begin{minipage}{0.82\textwidth}
\footnotesize
\textit{Notes}: The null hypotheses are formulated in \eqref{Eq::Test_final}. The results concern the subsample of married couples, versus married+cohabiting in the baseline.
\end{minipage}
\end{center}
\end{table}
\clearpage

\begin{table}[t]  
\begin{center}
\caption{Commitment test $p$-values, couples with children}\label{table::test_wchildren}
\begin{tabular}{L{2.5cm}C{2cm}C{2cm}C{2cm}C{2cm}}
\toprule
           &                          &                          & \multicolumn{2}{c}{new hypotheses}\\
\cmidrule{4-5} 
           & ${\cal H}_{0}^\text{FC}$ & ${\cal H}_{0}^\text{NC}$ & ${\cal H}_{0}^\text{NC-new}$             & ${\cal H}_{0}^\text{LC}$   \\ 
\midrule
Austria    	&  	$<0.001$  	&  	$<0.001$  	&  	$<0.001$  	&  	$0.806$  	\\
Belgium    	&  	$0.073$  	&  	$0.074$  	&  	$0.047$  	&  	$0.811$  	\\
Bulgaria   	&  	$0.166$  	&  	$0.126$  	&  	$0.193$  	&  	$0.292$  	\\
Cyprus     	&  	$0.471$  	&  	$0.484$  	&  	$0.598$  	&  	$0.877$  	\\
Czech Rep. 	&  	$0.015$  	&  	$0.854$  	&  	$0.882$  	&  	$0.829$  	\\
France     	&  	$0.011$  	&  	$0.021$  	&  	$0.038$  	&  	$0.330$  	\\
Greece     	&  	$0.002$  	&  	$<0.001$  	&  	$<0.001$  	&  	$0.045$  	\\
Hungary    	&  	$0.048$  	&  	$0.022$  	&  	$0.033$  	&  	$0.906$  	\\
Italy      		&  	$0.003$  	&  	$<0.001$  	&  	$0.001$  	&  	$0.320$  	\\
Luxembourg &  	$0.021$  	&  	$0.010$  	&  	$0.013$  	&  	$0.720$  	\\
Poland     	&  	$0.004$  	&  	$0.004$  	&  	$0.009$  	&  	$0.717$  	\\
Romania    	&  	$<0.001$  	&  	$<0.001$  	&  	$<0.001$  	&  	$0.441$  	\\
Slovakia   	&  	$0.117$  	&  	$0.341$  	&  	$0.242$  	&  	$0.398$  	\\
Spain      	&  	$0.594$  	&  	$0.386$  	&  	$0.484$  	&  	$0.746$  	\\
UK         		&  	$0.255$  	&  	$0.260$  	&  	$0.360$  	&  	$0.222$  	\\
\bottomrule
\end{tabular}
\begin{minipage}{0.82\textwidth}
\footnotesize
\textit{Notes}: The null hypotheses are formulated in \eqref{Eq::Test_final}. The results concern the subsample of couples with children, versus couples with and without children in the baseline.
\end{minipage}
\end{center}
\end{table}
\clearpage

\begin{table}[t]  
\begin{center}
\caption{Commitment test $p$-values, couples aged 30-60}\label{table::test_3060}
\begin{tabular}{L{2.5cm}C{2cm}C{2cm}C{2cm}C{2cm}}
\toprule
           &                          &                          & \multicolumn{2}{c}{new hypotheses}\\
\cmidrule{4-5} 
           & ${\cal H}_{0}^\text{FC}$ & ${\cal H}_{0}^\text{NC}$ & ${\cal H}_{0}^\text{NC-new}$             & ${\cal H}_{0}^\text{LC}$   \\ 
\midrule
Austria    	&  	$<0.001$  	&  	$<0.001$  	&  	$<0.001$  	&  	$0.848$  	\\
Belgium    	&  	$0.058$  	&  	$0.154$  	&  	$0.037$  	&  	$0.477$  	\\
Bulgaria   	&  	$0.096$  	&  	$0.081$  	&  	$0.131$  	&  	$0.354$  	\\
Cyprus     	&  	$0.248$  	&  	$0.315$  	&  	$0.422$  	&  	$0.809$  	\\
Czech Rep. 	&  	$<0.001$  	&  	$0.056$  	&  	$0.094$  	&  	$0.770$  	\\
France     	&  	$0.003$  	&  	$0.007$  	&  	$0.013$  	&  	$0.249$  	\\
Greece     	&  	$0.003$  	&  	$0.002$  	&  	$0.002$  	&  	$0.234$  	\\
Hungary    	&  	$0.086$  	&  	$0.055$  	&  	$0.093$  	&  	$0.999$  	\\
Italy      		&  	$<0.001$  	&  	$<0.001$  	&  	$<0.001$  	&  	$0.403$  	\\
Luxembourg &  	$0.003$  	&  	$<0.001$  	&  	$0.002$  	&  	$0.661$  	\\
Poland     	&  	$0.008$  	&  	$0.004$  	&  	$0.008$  	&  	$0.798$  	\\
Romania    	&  	$<0.001$  	&  	$0.002$  	&  	$0.005$  	&  	$0.378$  	\\
Slovakia   	&  	$0.056$  	&  	$0.257$  	&  	$0.081$  	&  	$0.410$  	\\
Spain      	&  	$0.107$  	&  	$0.058$  	&  	$0.080$  	&  	$0.346$  	\\
UK         	&  		$0.370$  	&  	$0.405$  	&  	$0.519$  	&  	$0.327$  	\\
\bottomrule
\end{tabular}
\begin{minipage}{0.82\textwidth}
\footnotesize
\textit{Notes}: The null hypotheses are formulated in \eqref{Eq::Test_final}. The results concern the subsample of couples aged 30-60, versus couples aged 21-65 in the baseline.
\end{minipage}
\end{center}
\end{table}
\clearpage

\begin{table}[t]  
\begin{center}
\caption{Commitment test $p$-values, extended test with coefficients on $t-2$ shocks}\label{table::test_historical}
\begin{tabular}{L{2.5cm}C{2cm}C{2cm}C{2cm}C{2cm}}
\toprule
           &                          &                          & \multicolumn{2}{c}{new hypotheses}\\
\cmidrule{4-5} 
           & ${\cal H}_{0}^\text{FC}$ & ${\cal H}_{0}^\text{NC}$ & ${\cal H}_{0}^\text{NC-new}$             & ${\cal H}_{0}^\text{LC}$   \\ 
\midrule
Austria    	&  	$0.005$  	&  	$0.005$  	&  	$0.003$  	&  	$0.835$  	\\
Belgium    	&  	$<0.001$  	&  	$<0.001$  	&  	$<0.001$  	&  	$0.547$  	\\
Bulgaria   	&  	$0.487$  	&  	$0.341$  	&  	$0.426$  	&  	$0.247$  	\\
Cyprus     	&  	$0.005$  	&  	$0.002$  	&  	$0.003$  	&  	$0.004$  	\\
Czech Rep. 	&  	$<0.001$  	&  	$0.002$  	&  	$0.004$  	&  	$0.562$  	\\
France     	&  	$<0.001$  	&  	$<0.001$  	&  	$<0.001$  	&  	$0.846$  	\\
Greece     	&  	$0.025$  	&  	$0.010$  	&  	$0.017$  	&  	$0.980$  	\\
Hungary    	&  	$0.245$  	&  	$0.701$  	&  	$0.774$  	&  	$0.851$  	\\
Italy      		&  	$<0.001$  	&  	$<0.001$  	&  	$<0.001$  	&  	$0.927$  	\\
Luxembourg &  	$0.004$  	&  	$0.008$  	&  	$0.006$  	&  	$0.836$  	\\
Poland     	&  	$0.518$  	&  	$0.358$  	&  	$0.432$  	&  	$0.715$  	\\
Romania    	&  	$0.051$  	&  	$0.140$  	&  	$0.195$  	&  	$0.462$  	\\
Slovakia   	&  	$0.158$  	&  	$0.262$  	&  	$0.328$  	&  	$0.602$  	\\
Spain      	&  	$0.470$  	&  	$0.534$  	&  	$0.621$  	&  	$0.538$  	\\
UK         	&  		$0.808$  	&  	$0.668$  	&  	$0.736$  	&  	$0.332$  	\\
\bottomrule
\end{tabular}
\begin{minipage}{0.82\textwidth}
\footnotesize
\textit{Notes}: The null hypotheses are based on \eqref{Eq::Test_final} but extended with restrictions on the coefficients on wage shocks at $t-2$. They are given, for $j\in\{ 1,2 \}$ and $\tau \in \{1,2\}$, by
\begin{equation*}
\begin{array}{rllll}
{\cal H}_{0}^\text{FC}:    ~    &\beta_{j[\omega_{-jt}]}=0,   & \beta_{j[\omega_{jt-\tau}]}=0, & \beta_{j[\omega_{-jt-\tau}]}=0, & \beta_{j[\theta_j]}=0; \\
{\cal H}_{0}^\text{NC}:    ~    &                             & \beta_{j[\omega_{jt-\tau}]}=0, & \beta_{j[\omega_{-jt-\tau}]}=0, & \beta_{j[\theta_j]}=0; \\
{\cal H}_{0}^\text{NC-new}: ~   & \beta_{j[\omega_{-jt}]}>0,  & \beta_{j[\omega_{jt-\tau}]}=0, & \beta_{j[\omega_{-jt-\tau}]}=0, & \beta_{j[\theta_j]}=0; \\
{\cal H}_{0}^\text{LC}: ~       & \beta_{j[\omega_{-jt}]}>0,  & \beta_{j[\omega_{jt-\tau}]}<0, & \beta_{j[\omega_{-jt-\tau}]}>0, & \beta_{j[\theta_j]}<0.
\end{array}
\end{equation*}
\end{minipage}
\end{center}
\end{table}
\clearpage

\end{document}